\documentclass[
reprint,
superscriptaddress,
amsmath,amssymb,amsfonts,aps,
longbibliography,
prb,
]{revtex4-2}
\usepackage[T1]{fontenc}
\usepackage[utf8]{inputenc}
\usepackage{amstext}

\usepackage{latexsym,epsfig,graphicx}
\usepackage{dcolumn}
\usepackage{subfigure}
\usepackage{comment}
\usepackage{color}
\usepackage{bm}
\usepackage{mathrsfs}
\usepackage{tensor}
\usepackage{xspace, xcolor}
\usepackage{epstopdf}
\usepackage{tabularx}
\usepackage{longtable}
\usepackage{lipsum} 
\usepackage[colorlinks=true, pdfstartview=FitV, linkcolor=red, citecolor=blue, urlcolor=blue]{hyperref}
\usepackage[normalem]{ulem}
\usepackage{natbib}
\usepackage[version=4]{mhchem}
\usepackage{braket}
\usepackage{url}

\newcommand{\dd}{\mathrm{d}}
\newcommand{\lp}{\left( }
\newcommand{\rp}{\right) }

\makeatother

\begin{document}
\title{Emergence of Gibbs ensembles as steady states in Lindbladian dynamics}
\author{Shi-Kang Sun}
\affiliation{Beijing National Laboratory for Condensed Matter Physics, Institute of Physics, Chinese Academy of Sciences, Beijing 100190, China}
\affiliation{School of Physical Sciences, University of Chinese Academy of Sciences, Beijing 100049, China}
\author{Shu Chen}
\email{schen@iphy.ac.cn }
\affiliation{Beijing National Laboratory for Condensed Matter Physics, Institute of Physics, Chinese Academy of Sciences, Beijing 100190, China}
\affiliation{School of Physical Sciences, University of Chinese Academy of Sciences, Beijing 100049, China}
\date{\today}

\begin{abstract}
We explicitly construct unique non-equilibrium steady state (NESS) of Lindblad master equation characterized by a Gibbs ensemble  $\rho_{\text{NESS}} \propto e^{-\beta \tilde{H}}$, where the effective hamiltonian $\tilde{H}$ is an element in the center of the commutant algebra $\mathcal{C}$ of the original hamiltonian. Specifically, if $\mathcal{C}$ is Abelian, then $\tilde{H}$ consists only of $U(1)$ conserved charges of the original Hamiltonian. When the original Hamiltonian has multiple charges, it is possible to couple them with bathes at different temperature respectively, but still leads to an equilibrium state. Multiple steady states arise if the number of bathes is less than the number of charges. To access the Gibbs NESS, the jump operators need to be properly chosen to fulfill quantum detailed balance condition (qDBC). These jump operators are ladder operators for $\tilde{H}$ and jump process they generate form a vertex-weighted directed acyclic graph (wDAG). By studying the XX model and Fredkin model, we showcase how Gibbs states emerge as steady states.
\end{abstract}

\maketitle

\section{Introduction}
The Lindblad master equation is a paradigm for describing Markovian open quantum systems \citep{10.1063/1.522979GKS, Lindblad1976, oqsbook}. It represents the dynamics as an average of different trajectories, which consist of non-Hermitian evolution and stochastic quantum jumps, also known as "unraveling." In master equation form, the dynamics generator is
\begin{equation}\label{eq:lindblad}
	\mathcal{L}(\rho) = -i  H_{\text{eff}} \rho + i \rho H_{\text{eff}}^{\dagger}  + \sum_{k} L_k \rho L_k^{\dagger}  \, .
\end{equation}
where $L_k$s are jump operators and $H_{\text{eff}} = H - \frac{i}{2}\sum_k L_k^{\dagger}L_k$. The detailed balance condition (DBC) is not required in Lindblad dynamics, making it powerful and capable of characterizing non-equilibrium phenomena \citep{Kamenev_2011, Derrida_2007, GARRAHAN2018130, Prosen_2008,Verstraete2009, PhysRevResearch.5.L012003, PhysRevLett.132.216301, buvca2019non, rakovszky2023defining, wang2023superdiffusive, PhysRevLett.127.070402, PhysRevLett.129.250601, PhysRevB.106.064203, PhysRevB.108.024404}. Recently with development of quantum computation, people regain interests in crafting quantum Gibbs samplers and preparing specific state utilizing Lindblad dynamics \citep{Rall2023thermalstate, chen2023quantum, ding2024efficient, guo2024designing}. In this work we make direct use of qDBC to derive exact Gibbs state that only depends on $U(1)$ conserved charges of original Hamiltonian. More generally, the state is in the center of the commutant algebra \citep{moudgalya2022hilbert} $\mathcal{C}$ of the original hamiltonian.

DBC originates from classical statistical physics and guarantees an equilibrium stationary state. In a classical Markov process generated by W, the probability distribution of configuration set \{c\} satisfies
$
	\frac{\dd P(c)}{\dd t} = \sum_{c^{\prime} \neq c} W(c^{\prime} \rightarrow c)P(c^{\prime}) - W(c\rightarrow c^{\prime}) P(c)
$.
For a stationary state satisfying classical DBC, $W(c^{\prime} \rightarrow c)P(c^{\prime}) = W(c\rightarrow c^{\prime}) P(c)$. DBC can also be interpreted as time-reversal symmetry, as it constrains the relative probabilities of configurations connected by a Markov process. In other words, starting from an arbitrary configuration, one can determine the probability of any other configuration without frustration. Besides the above intuitive condition, Kubo-Martin-Schwinger (KMS) condition \citep{Haag1967} $\langle \hat{A}(t)\hat{B}(0)\rangle_{\text{eq}} = \langle \hat{B}(-t-i \beta)\hat{A}(0)\rangle_{\text{eq}}$ also acts as definition for thermal equilibrium state. It suggests a generalized "time evolution" acting on $\hat{B}$. qDBC extends this idea by requiring that this "time evolution" is also a completely positive and trace-preserving (CPTP) map, or a quantum Markov semigroup (QMS). Similar idea is applied to study properties of thermal states \citep{10.1063/5.0167353} and mixing time \citep{PhysRevLett.130.060401}, to solve some models exactly \citep{PhysRevB.109.054311, PRXQuantum.2.020336, PhysRevLett.131.190403ExactSolution} and a local thermal state is constructed \citep{PhysRevE.100.022111}. In the context of Lindblad dynamics $\mathcal{L}(H, \{L_k\})$, the qDBC condition for an invertible (faithful) density matrix $\rho$ manifests as \citep{ALICKI1976249, qdbc}
\begin{equation}\label{eq:qDBC}
	\left[ H, \rho \right]  = 0, ~ \text{and} ~
	\tilde{L}_k \rho = \lambda_k \rho \tilde{L}_k \, , \forall k, \, \lambda_k >0
\end{equation}
where $\tilde{L}_k=U L_k$, $U$ is an unitary that operates on space spanned by \{$L_k$\}. Liouvillian of this form that hosts a Gibbs NESS is often denoted as Davies generator \citep{Davies1974}.

When jump operators are Hermitian, the identity matrix is always a NESS due to hermiticity of the Liouvillian. However, when jump operators are non-Hermitian and the Liouvillian is not diagonal-preserving, the NESS is typically not a diagonal equilibrium state and usually has finite stationary current, especially in the case of boundary driving \citep{Prosen_2015}. Nevertheless, in this work we demonstrate that a diagonal equilibrium state can be constructed as a NESS in various models. Such a state satisfies qDBC, making it an equilibrium state that depends only on the conserved quantities of the original Hamiltonian. 

Our method introduced here relies on the coherent Hamiltonian having conservation laws, such that its Hilbert space is fractured into some disconnected pieces, which we denote as Krylov sectors or dynamical sectors. Besides conventional $U(1)$ conversational laws like particle conservation \citep{Buca_2012}, it is also applicable to non-Abelian cases \citep{Zhang_2020},  dynamical constrained systems \citep{doi:10.1080/0001873031000093582, PhysRevLett.53.1244} and systems with Hilbert space fragmentation (HSF) \citep{PhysRevX.10.011047dipole, moudgalya2022hilbert, PhysRevResearch.5.043239hsfopen,  PhysRevX.14.021034word}.

\section{General theory}
Throughout this work we only consider separable Hilbert space with finite local dimension, such as spin system or fermion system.
Our goal is to achieve a hierarchical structure of NESS that resembles equilibrium Gibbs distribution. Since a density matrix $\rho$ is positive and Hermitian, it can be expressed as $\rho=e^{-\beta \tilde{H}}$, where $\tilde{H}$ is a Hermitian matrix and $\beta \in \mathbb{R}$. To clarify Gibbs state, we construct $\tilde{H}$ by conserved quantities in original Hamiltonian and relate $\beta$ to temperature-alike quantity as in classical statistical physics.
To realize this hierarchy, we pack states by taking conserved quantities as our packing principle. Since conserved quantities separate Hilbert space into disconnected sectors, every state sharing the same quantum numbers should have the same weight in $\rho$.  An effective tool to characterize these conservation laws is commutant algebra. The commutant algebra formalism, as discussed in \citep{moudgalya2022hilbert}, is particularly useful for analyzing HSF, but in general it is not restricted to HSF cases. Commutant algebra formalism regards every invariant subspace of Hamiltonian as representation space for each irreducible representation (irrep) of bond algebra $\mathcal{A}$ generated by Hamiltonian terms $\{h_{\text{loc}}\}$ and identity operator $I$. The commutant algebra $\mathcal{C}$ consists of all operators $\{O\}$ of Hilbert space $\mathcal{H}$ that commute with all generators of $\mathcal{A}$, i.e., $[h_{\text{loc}, i}, O]=0$ for all $i$. The total Hilbert space can be decomposed by irreps of bond algebra and commutant algebra marked by $\lambda$, i.e.,
$
	\mathcal{H} = \bigoplus_{\lambda} \mathcal{H}^{\mathcal{A}}_{\lambda}\otimes \mathcal{H}^{\mathcal{C}}_{\lambda}
$.
If $\mathcal{H}^{\mathcal{C}}_{\lambda}$ has dimension $d$ larger than 1, then the Hamiltonian has degenerated sectors $H_{\lambda, 1,...d}$. An operator in $\mathcal{A}$ has form $\oplus_{\lambda}h_{\lambda}\otimes \mathcal{I}$, and an operator in $\mathcal{C}$ has form $\oplus_{\lambda}\mathcal{I}\otimes c_{\lambda}$ where $h_{\lambda}$ and $c_{\lambda}$ is supported on $\mathcal{H}^{\mathcal{A}}_{\lambda}$ and $\mathcal{H}^{\mathcal{C}}_{\lambda}$, respectively.

To pack states with the same quantum numbers, it is straightforward to consider the dynamical sector. Every dynamical sector corresponds to an invariant subspace $\mathcal{H}^{\mathcal{A}}_{\lambda, a}$ under the application of Hamiltonian, where $a$ is the degeneracy index. Denote a dynamical sector with representative state $\ket{\omega}$ as $\psi(\ket{\omega})$. By definition, for state $\ket{\omega^{\prime}_{\psi^{\prime}}}$ and $\ket{\omega_{\psi}}$ in different sector, $\bra{\omega^{\prime}_{\psi^{\prime}}}H^k\ket{\omega_{\psi}}=0$ for all $k$. The totally mixed state $\rho_{\psi}$ of sector $\psi(\ket{\omega})$, which is proportional to the projector to this subspace, commutes with Hamiltonian, i.e.,
\begin{equation}\label{eq:ansatz}
	[H, \rho_{\psi}] = 0, \, \rho_{\psi} = \frac{1}{D_{\psi}}\sum_{i\subset \psi(\ket{\omega})}\ket{i}\bra{i} = P_{\psi} \, ,
\end{equation}
where $D_{\psi}$ is the dimension of this dynamical sector and we have omitted $\omega$.

If commutant algebra is non-Abelian, some dynamical sectors are degenerate so it becomes
\begin{equation}
		[H, \rho_{\psi}] = 0, \, \rho_{\psi} = \frac{1}{D_{\psi}\,  \text{Tr}(M_{\psi})}\sum_{i\subset \psi(\ket{\omega})}\ket{i}\bra{i} \otimes M_{\psi} \, ,
\end{equation}
where $M_{\psi}$ is a Hermitian matrix with dimension equal to the degeneracy of $\psi$. In this work we only consider Abelian commutant algebra and leave the non-Abelian case for future work.

Next we anticipate a Gibbs state ansatz for NESS: it consists of different $\rho$ with specific weights $c_{\psi}$, i.e.,
\begin{equation}\label{eq:ness}
	\rho_{\text{NESS}} \propto \sum_{\psi} c_{\psi} \rho_{\psi} \, .
\end{equation}
To make sure the above ansatz holding true, we should assure jump operators to be chosen properly to keep consistent with the ansatz. Under this construction $\rho$ automatically satisfies $[H, \rho_{\text{NESS}}]=0$ and is in commutant algebra $\mathcal{C}$ of Hamiltonian.
To guarantee the validity of the ansatz, we further require that $\rho$ is the zero eigenstate of dissipation
$
	\mathcal{D}(\rho) = 0$ where $\mathcal{D}(\cdot) = \sum_{L_i \subset \text{Dis}}  \lp  L_i \cdot L_i^{\dagger} - \frac{1}{2} \{L_i^{\dagger}L_i, \cdot \} \rp
$.
By implementing qDBC (\ref{eq:qDBC}), it can be automatically satisfied if jump operators include both operator $\hat{O}$ and its adjoint operator $\hat{O}^{\dagger}$, so it is indeed a NESS.

 Next let us examine how the coefficient $c_{\psi}$ is fixed. Since $c_{\psi}$ for any state $\ket{i}\bra{i}$ in $\rho_{\psi}$ is same, we can focus on one arbitrary state. In the following we will work in computational basis. It is useful to gain insights from a simple example. For simplicity suppose that the Hamiltonian conserves magnetization and jump operators are just $\sigma^{\pm}$ acting on some sites with strength $\gamma_1$ and $\gamma_2$. After applying a single jump operator to a pure state $\ket{i}\bra{i}$ with weight $c_{\psi}$, we may get either $\mathcal{D}_{L}(c_{\psi}\ket{i}\bra{i}) = 0$ or, for example, $c_{\psi}\gamma_1 \lp\ket{i^{\prime}}\bra{i^{\prime}}-\ket{i}\bra{i}\rp$, where $\ket{i^{\prime}}\bra{i^{\prime}}$ is in a different sector $\psi^{\prime}$. Note that we also have reverse process but with a different rate $\gamma_2$ as $\mathcal{D}_{L^{\dagger}}(c_{\psi^{\prime}}\ket{i^{\prime}}\bra{i^{\prime}})=c_{\psi^{\prime}}\gamma_2 \lp\ket{i}\bra{i}-\ket{i^{\prime}}\bra{i^{\prime}}\rp$. In the spirit of DBC, we must have $\frac{c_{\psi}}{c_{\psi^{\prime}}}=\frac{\gamma_2}{\gamma_1} \equiv e^{-\beta}$. Thus if we have fix $c_{\psi}$, we can get $c_{\psi^{\prime}}$. Repeating this process, we can get all $c_{\psi}$, given that there is no any contradictory. In the end we get a NESS of Gibbs form
\begin{equation}\label{eq:expform}
	\rho_{\text{NESS}}(\psi) \propto e^{-\beta \tilde{H}(\psi)},
\end{equation}
where $\tilde{H}(\psi)$ is an effective modular Hamiltonian of the equivalent class $\psi$. Since $\beta$ is related  to dissipation strength, it can effectively characterize "temperature".

Now returning to qDBC (\ref{eq:qDBC}), natural questions arise: how is $\tilde{H}$ related to conserved quantities of the original Hamiltonian and how can we find specific jump operators that satisfies qDBC? We will see they are related. First let us consider choice of jump operators. If $\lambda_k=1$, then a direct choice for $\tilde{L}_k$ to satisfy qDBC will be an element of bond algebra or a diagonal operator like $S^z$ in all irreps, for example, see  \citep{PhysRevResearch.5.043239hsfopen}. But this choice for jump operators preserves (parts of) dynamical sector structure, so there will be a NESS in each sector. In this way we generally get an initial state dependent NESS. If there is weak symmetry $P$ that $\mathcal{L}(P\rho P)=P\rho P$ or non-trivial strong symmetry \citep{PhysRevLett.128.033602, Buca_2012, PhysRevA.89.022118, Zhang_2020, li2024highlyentangled} $S$ that commutes with all $\tilde{L}$ and $H$ (i.e., $S$ is in commutant of algebra of $\langle \{L_k\}, \{L^{\dagger}_k\}, H \rangle$, $[S, L_k]=[S,L^{\dagger}_k]=[S,H]=0$ for all $k$), we will have multiple NESS, because the Liouvillian is reducible \citep{evans1977irreducible, Frigerio1977, Baumgartner_2008,Zhang_2024} to independent sectors determined by invariant space of $S$.

A more interesting case, which is the focus of this work, is that $\tilde{L}_k$ is neither in bond algebra and commutant algebra nor in diagonal form. This choice for $\tilde{L}_k$ tends to break fragmentation structure and is a necessary condition for uniqueness of the NESS. We define the centralizer of $\rho$ as $C_{\rho}$. If $C_{\rho} \subseteq \lp \mathcal{A} \cup \mathcal{C}\rp$, then NESS is not unique because each sector evolves independently and thus retains initial state information.

For the case $\lambda_k \neq 1$, we can regard $\tilde{L}_k$ as a dynamical symmetry. Given (\ref{eq:qDBC}) and (\ref{eq:expform}), defining $\text{Ad}_X(Y) \equiv XYX^{-1}$ and $\text{ad}_{X}(Y) \equiv [X, Y]$, we have
\begin{equation}
	e^{\beta \tilde{H}} \tilde{L}_k e^{-\beta \tilde{H}} = \text{Ad}_{e^{\beta \tilde{H}}}(\tilde{L}_k) = e^{\text{ad}_{\beta \tilde{H}}} (\tilde{L}_k)= \lambda_k \tilde{L}_k \, ,
\end{equation}
which is satisfied by $[\tilde{H}, \tilde{L}_k]=\frac{\log(\lambda_k)}{\beta} \tilde{L}_k$. So $\tilde{L}$s are ladder operators for $\tilde{H}$. We can thus take $\tilde{H}$ to be conserved quantities of the original Hamiltonian. Denote $\varepsilon = \frac{\log(\lambda_k)}{\beta}$ and we have a tower of states of $\tilde{H}$ with energy $\{E_0, E_0+\varepsilon, E_0+2\varepsilon, ... \}$ and corresponding states are $\{\ket{\phi}, \tilde{L}_k \ket{\phi}, \tilde{L}_k^2 \ket{\phi},... \}$ for $\tilde{H}\ket{\phi}=E_0 \ket{\phi}$ and etc.. To define "temperature" we may let $\beta = \log(\lambda_k)$ so $\varepsilon = 1$. The $\tilde{H}$ behaves as Hamming distance from the root state $\ket{\phi}$. Since all states in the same sector $\psi(\ket{\phi})$ has identical weight in NESS (\ref{eq:ness}), there must be some additional constraints on jump operators. We can think of a dynamical sector as one equivalent class state $|\psi(\phi))$ and jump operators act as transitions between these states, $\tilde{L}|\psi(\phi)) \equiv \tilde{L} \rho_{\psi(\phi)} \tilde{L}^{\dagger}$ and $(\psi(\phi)|\psi(\omega)) =  \text{Tr}(\rho_{\psi(\phi)}^{\dagger}\rho_{\psi(\omega)})$. Viewing dynamical sectors as vertices of a graph, then each edge corresponds to an application of jump operator $\tilde{L}$ and carries a positive weight $e^{-\beta}$ according to proportion of dissipation strength of $\tilde{L}$ and $\tilde{L}^{\dagger}$. The weight of a path is then defined as the product of all weights on the path and is related to energy of $\tilde{H}$. A key point in determining all $c_{\psi}$, or validity of the ansatz (\ref{eq:ness}), relies on the fact that the graph being a weighted directed acyclic graph (wDAG) generated by a set of $\{\tilde{L}_i\}$, and all path with the same source and end must have the same weight, which implies jump operators that carries different weights $\beta_a$, $\beta_b$ cannot generate the same path $(\psi(\omega)|\tilde{L}_i^{a\dagger} e^{i H \epsilon} \tilde{L}_j^b  |\psi(\omega)) = (\psi(\omega)|\tilde{L}_i^{a}  e^{i H \epsilon} \tilde{L}_j^b  |\psi(\omega))=0$ for arbitrary $\epsilon$, $\psi(\omega),i,j$ if $a\neq b$.

More generally, we can choose $\tilde{H}$ to be an element of Cartan subalgebra of the commutant algebra $\mathcal{C}$. A Cartan subalgebra is the maximal Abelian subalgebra generated by commuting elements of an algebra. Then $\tilde{L}_k$s are ladder operators like what we have in Cartan-Weyl method for constructing irrep basis for a Lie algebra. In this way we get a tower of states consisting states of different dynamical sector. However we should take care as the commutant algebra need not be a Lie algebra, and thus $\tilde{L}_k$ need not to be in commutant algebra $\mathcal{C}$. For $\mathcal{C}$ is Abelian, as in common particle-conserving or classical fragmentation case, $\tilde{L}_k$s naturally lie out of $\mathcal{C}$. For $\mathcal{C}$ is non-Abelian, we need at least two kinds of $\tilde{L}_k$: one is in $\mathcal{C}$ that connects sectors within a $\mathcal{H}_{\lambda}^{\mathcal{C}}$ and another out of $\mathcal{C}$ that connects different $\mathcal{H}_{\lambda}^{\mathcal{C}}$. For example, if $\mathcal{C} = \mathfrak{su}(2)$ every sector is labeled by $\ket{J,j=-J,-J+1,...,J}$. We can choose $\tilde{L}_1 =\sum_i \sigma^{+}_i$ to create a thermal state in every $J$ sector. However it is difficult to write an explicit operator to relate different $J$ sector. In two-qubit case since we have only triplets and singlet, one choice of $\tilde{L}_2$ can be $\tilde{L}_2= \lp \sigma^{+}_1 + \sigma^{+}_2\rp\lp I-\sigma^z_1 \sigma^z_2\rp$. After resolving degeneracy between $\mathcal{H}_{\lambda}^{\mathcal{A}}$ within a $\mathcal{H}_{\lambda}^{\mathcal{C}}$, we come to the same situation as in Abelian commutant algebra. So in the following we only consider jumps between different sectors in Abelian case.

\begin{figure}[tb]
	\centering
	\includegraphics[width=0.9\linewidth]{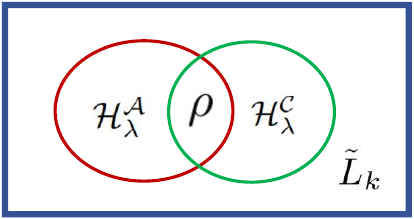}
	\caption{Illustration of the steady state $\rho$, bond algebra $\mathcal{A}$, commutant algebra $\mathcal{C}$ and jump operators $\tilde{L}_k$ in the operator space.}
	\label{fig:ss}
\end{figure}

In summary, we conclude that the steady state is in the center of the commutant algebra $\mathcal{C}$ of the original hamiltonian $H$. The state is by definition in $\mathcal{C}$, and it is also in the bond algebra $\mathcal{A}$ since it is a sum of identities and commutes with every element of $\mathcal{C}$. The double-commutant theorem assures the intersection of $\mathcal{A}$ and $\mathcal{C}$ is the center for both of them \citep{landsman1998lecturenotescalgebrashilbert}. The relevant jump operators lie outside $\mathcal{A}$ and $\mathcal{C}$ for the uniqueness of the steady state, as shown in Fig.~\ref{fig:ss}.

\section{Equilibrium environment}
To gain an intuitive view, let us consider the canonical XX model
\begin{equation}
	H_{XX} = \sum_i \sigma_i^x \sigma_{i+1}^x + \sigma_i^y \sigma_{i+1}^y \, .
\end{equation}
in contact of equilibrium environment with the jump operators given by pairs of $\sigma^{\pm}$ acting on the i-th site
$
	L_\text{Dis} = \{L_{i,+}\equiv \sqrt{\gamma_1}\sigma_{i}^{+} ,\, L_{i,-}\equiv \sqrt{\gamma_2} \sigma_{i}^{-} \}
$.
Since they corresponds to raising and lowering of total magnetization, we know
$\rho_{\text{NESS}}\propto e^{-\beta \tilde{H}}$ with $\beta = \log(\frac{\gamma_1}{\gamma_2})$ and the effective Hamiltonian is just the total magnetization $\tilde{H} = -\sum_{i} \sigma^z_i \equiv -\sigma^z_{\text{tot}}$ \citep{PhysRevE.83.011108}.

Actually pairs of jump operators can act on multiple sites simultaneously, but the ratio between coefficient of $\sigma^{\pm}$ must be identical to ensure Eq.(\ref{eq:ness}) holding true. This means that only one $\beta$ is permitted in order to get a NESS of Gibbs form, as there is only one $U(1)$ charge for the XX model. Conversely, if we have multiple $\beta$s, then one sector can gain different weight due to path that carries distinct $\beta$, and the consistence of ansatz (\ref{eq:ness}) is broken. For example, for the system with $N=2$ and jump operators acting on both sites with $\beta_1\neq \beta_2$,  starting from sector $\psi(\omega=\ket{\downarrow\downarrow})$, we can reach sector composing of states $\ket{\uparrow\downarrow}$ and $\ket{\downarrow\uparrow}$ with weight $e^{-\beta_1}$ or $e^{-\beta_2}$, depending on on which end jump operator is applied. So the ranges of jump operators overlap, $(1-\delta_{a,b})(\psi(\omega)|\tilde{L}_i^{a\dagger} e^{i H \epsilon} \tilde{L}_j^b  |\psi(\omega)) \propto\sin^2 (\epsilon) \neq0$.

By adding terms without breaking particle number conservation to the Hamiltonian, the above discussion can be directly applied, e.g., the NESS of XXZ model is the same as XX model. As details of original Hamiltonian is lost in NESS, relevant information is encoded in their different relaxation dynamics.

\section{Non-equilibrium environment}
Next we study Fredkin spin chain \citep{doi:10.1142/S0129055X17500313fredkinchain} with boundary driving.
The Fredkin spin chain is composed of local three-site spin interactions and described by
\begin{equation}\label{eq:fredkin}
	H_F = \sum_i P^{\uparrow}_i \ket{S}\bra{S}_{i+1,i+2}+\ket{S}\bra{S}_{i,i+1}P^{\downarrow}_{i+2}\, ,
\end{equation}
where $P^{\uparrow}=(1+Z)/2$ and $\ket{S}$ is singlet state $\frac{1}{\sqrt{2}}\lp \ket{\uparrow \downarrow}-\ket{\downarrow \uparrow} \rp$. Denoting $\ket{\uparrow}, \ket{\downarrow}$ as $\ket{(}, \ket{)}$ and referring them as left and right parenthesis, respectively, we can represent a dynamical sector according to its canonical form $(k, b, a)$. A canonical form state is of form ()()$\cdots$k times$\cdots$ ()))$\cdots$ a times$\cdots$)(($\cdots$b times$\cdots$(. This structure is preserved by Fredkin dynamics, which can be easily checked by noting that every single Hamiltonian term can only changes a parenthesis configuration according to: ())$\leftrightarrow$)() and ()( $\leftrightarrow$ (().

We explicitly consider the Lindblad master equation of form (\ref{eq:lindblad}) for system of length $N$ with jump operators $\text{Dis}=\{\sqrt{\gamma_1}L_1, \sqrt{\gamma_2}L_1^{\dagger},\sqrt{\gamma_3}L_N, \sqrt{\gamma_4}L_N^{\dagger} \}$ where $L_i = \sigma^{+}_i$ and we define $\beta_l \equiv \log(\frac{\gamma_1}{\gamma_2})$, $\beta_r \equiv \log(\frac{\gamma_3}{\gamma_4})$.
Here we focus on the case with  $\beta_l = -\beta_r = \beta$. Discussions on a general case is given in SM \cite{appendix_note}. Without loss of generality we set $\gamma_1\geq\gamma_2$. Under this assumption particles flow into system from the left boundary and leave from the right driven by boundary jump terms.

Next we will derive exact NESS for Fredkin model under boundary driving. First we study conserved quantities of Fredkin model with length $N$. Since every dynamical sector can be represented by its canonical form state labeled by tuple ($N^k$ (paired parenthesis), $N^b$ (unpaired left parenthesis), $N^a$ (unpaired right parenthesis)), so we conclude the commutant algebra of Fredkin model is Abelian and is generated by non-local number operators that count paired parenthesis and unpaired left, right parenthesis, respectively. Explicitly, we have
\begin{widetext}
	\begin{equation}\label{eq:pairedp}
		N^k = \sum_{i=1}^{N-1} \sum^{\lfloor \frac{N-1-i}{2} \rfloor }_{m=0} P^{\uparrow}_i P^{\downarrow}_{i+2m+1} \prod_{ j=0}^{m-1} \lp 1-P^{\uparrow}_i P^{\downarrow}_{i+2j+1}\rp \lp 1-P^{\uparrow}_{i+2(m-j)} P^{\downarrow}_{i+2m+1}\rp
	\end{equation}
\end{widetext}
and $N^b = \sum_{i=1}^N P^{\uparrow}_i - N^k $, $N^a = \sum_{i=1}^N P^{\downarrow}_i - N^k$
where $P^{\uparrow}\equiv \frac{1}{2}\lp 1+\sigma^z \rp$ and $P^{\downarrow}\equiv \frac{1}{2}\lp 1-\sigma^z \rp$.
Since $2N^k + N^b + N^a = N$, only two of them are linear independent, and there is no other U(1) conservation law. Hence Fredkin model is not a model of HSF because its sectors show only algebraic growth with $N$, instead of an exponential one.

\begin{figure}[tb]
	\centering
	\includegraphics[width=1.0\linewidth]{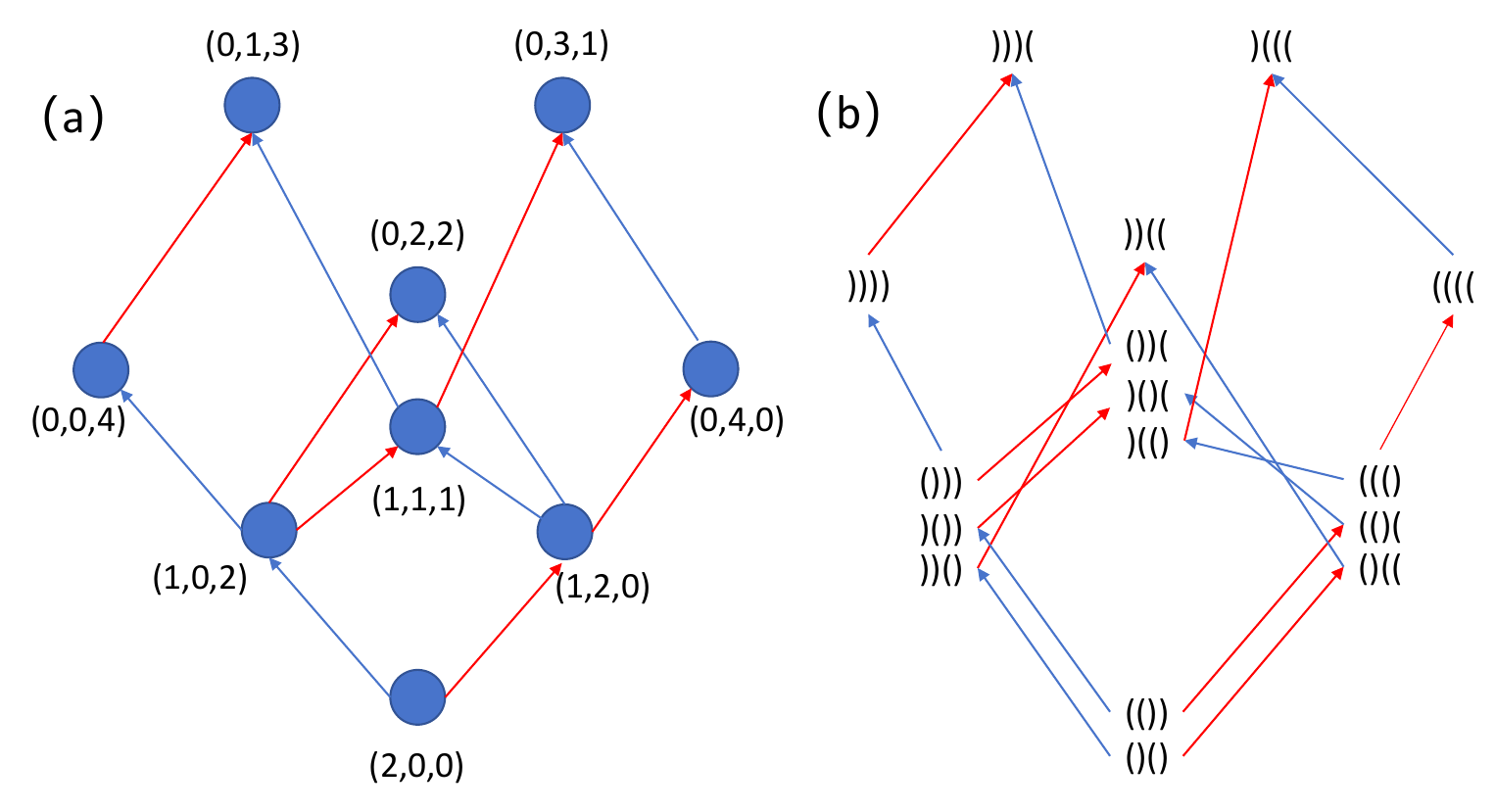}
	\caption{Structure of NESS of Fredkin spin chain with system length $N=4$. Tuples in (a) represent sector labeled by ($N^k$, $N^b$, $N^a$). Arrows pointing to upper left (blue) stand for application of $\sqrt{\gamma_2}L_1^{\dagger}$ and arrows pointing to upper right (red) stand for $\sqrt{\gamma_2}L_N$. The whole structure is generated from the root (2,0,0). Dots in (a) represent dynamical sectors and the corresponding states are shown in (b) with more details.}
	\label{fig:graph4}
\end{figure}

The $L_k$s we choose for Fredkin model do exactly raise and lower of these quantum numbers. A closer observation of Fredkin model tells that in Fredkin dynamics only paired parenthesis can move. An application of $L_{1,N}^{\dagger}$ on left (right) end of the chain breaks the pairing and creates two idle parenthesis. So $L_{1,N}^{\dagger}$ lowers $N^k$ by 1 and raises $N^{a,b}$ by 2, i.e., $[N^k, L_1^{\dagger}] = -L_1^{\dagger}$ and $[N^{a,b}, L_{1,N}^{\dagger}] = 2 L_{1,N}^{\dagger}$. If the jump operator $L^{\dagger}$ acts on bulk sites, for example on the second site, then it is not a ladder operator any more, because it can annihilate and also create a parenthesis pair, $L^{\dagger}\ket{(()}\propto\ket{())}$, and the ansatz (\ref{eq:ness}) is not applicable.
Note that for different state in the same dynamical sector, effect of $L_k$ can be different in changing quantum numbers. For example, we explicitly plot the hierarchical structure of NESS of Fredkin model for system with $N=4$ in Fig.\ref{fig:graph4}. One can travel to (0,2,2) and (1,1,1) with same weight starting from different state in (1,2,0) by applying $L_1^{\dagger}$. Since this wDAG structure is generated by repeated applications of $\sqrt{\gamma_2}L_1^{\dagger}$ and $\sqrt{\gamma_2}L_N$ starting from (2,0,0), we call the sector of (2,0,0) as the root of wDAG.

\begin{figure}[tb]
	\centering
	\includegraphics[width=1.0\linewidth]{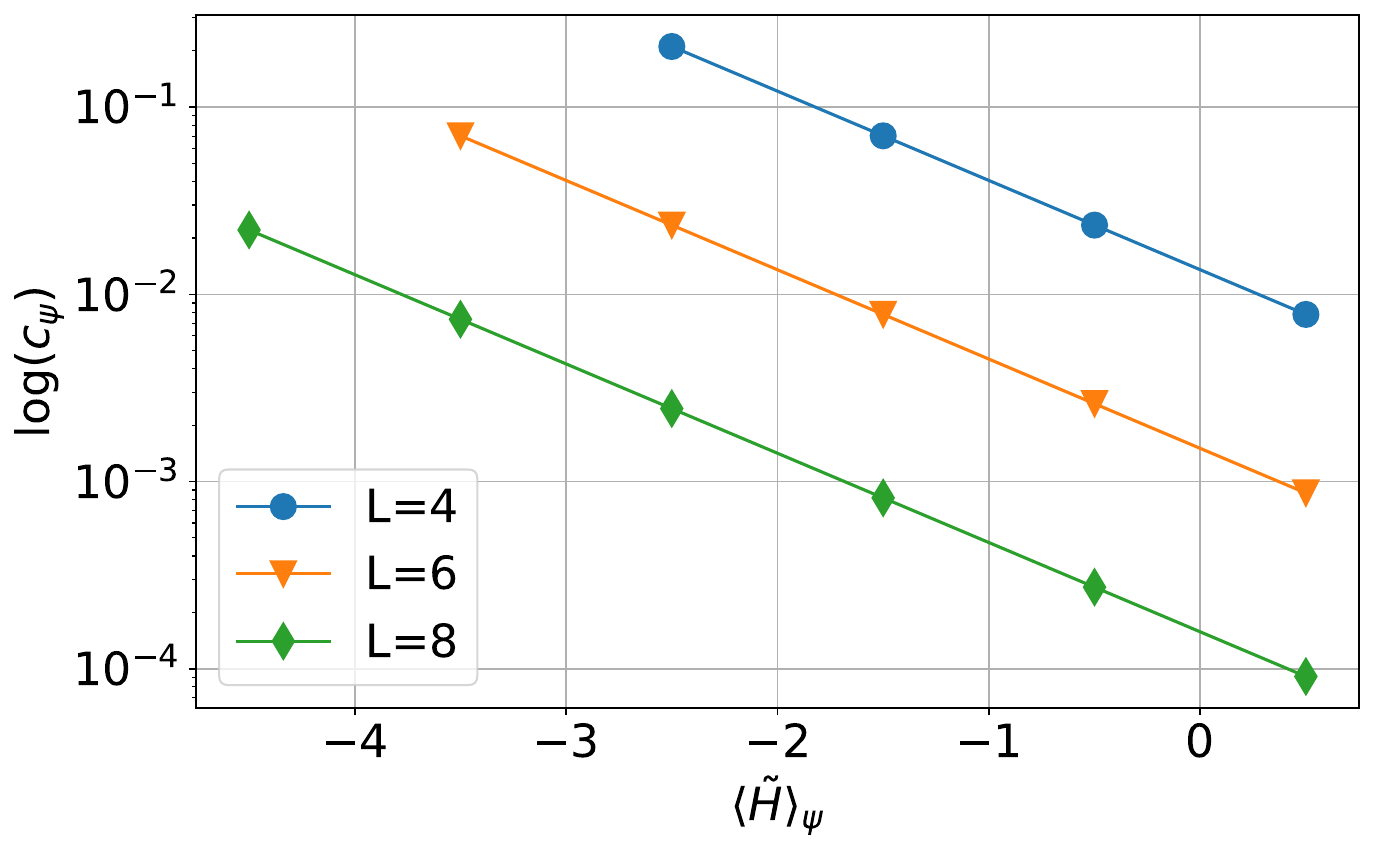}
	\caption{Logarithm of distinct coefficients $c_{\psi}$ in NESS of Fredkin spin chain under boundary driving with $\gamma_1 = 1.5$, $\gamma_2 = 0.5$ and N=4,6,8 versus the expectation value of $\tilde{H}$ in corresponding Krylov sector $\psi$. These coefficients satisfy perfect exponential relation, indicating that NESS is of form $\rho_{\text{NESS}}\propto e^{-\beta \tilde{H}}$.}
	\label{fig:logee}
\end{figure}

Then it follows that the NESS is $\rho\propto e^{-\beta \tilde{H}}$ with $\beta \equiv \log(\frac{\gamma_1}{\gamma_2})$ and
\begin{equation}\label{eq:Heff}
	\tilde{H} = -N^k - \frac{1}{4}\lp(-1)^{N^b} + (-1)^{N^a}\rp \, .
\end{equation}
This is a natural choice for $\beta$ because in $\tilde{H}$ there is no $\gamma$ dependent term. Difference between eigenvalues of $\tilde{H}$ can be interpreted as relative Hamming distance between sectors, e.g., it is proportional to the distance from $\psi (k,b,a)$ to another $\psi (k',b',a')$ in Fig.\ref{fig:graph4}. We numerically compute the NESS of this model. In computational basis the NESS is diagonal and its diagonal elements, which are just coefficients $c_{\psi}$ of basis states in Eq.(\ref{eq:ness}), are shown in Fig.\ref{fig:logee}, indicating a perfect exponential relation.

By "thermal equilibrium" we mean that the system is in equilibrium with bath, i.e., system exchanges particles or quantas of energy with bath according to DBC and NESS is a Gibbs state. By equilibrium statistical physics the "temperature" of system is naturally the same with bath. In terms of energy transport we can define temperature of bath $e^{\frac{1}{T_{\text{bath}}}}=\frac{\gamma_1}{\gamma_2}$ \citep{Derrida_2007}, where we choose the energy unit to be 1. However in boundary driven system the temperature is generally distinct at different boundary. 
The fact that Fredkin spin chain can be in thermal equilibrium under DBC with two baths at different temperatures is surprising. Actually it is general for systems with non-Abelian symmetry or multiple U(1) conserved charges, which makes coupling of different conserved charges with different baths possible. In this way it is also straight forward to have multiple steady states if the number of bathes is less than the number of charges. For example, in Fig.\ref{fig:graph4} deleting all red(blue) arrows will generate 3 independent Gibbs states.

Now we discuss two limits of $\beta\rightarrow 0 ,\, \infty $. In the infinite temperature case,  $\gamma_1 = \gamma_2$, so $L_k$s can be turned into Hermitian form. The NESS is an identity operator, representing an infinite temperature state as expected. In the zero temperature case, $\gamma_2 \rightarrow 0$ and NESS should relate to the ground state of $\tilde{H}$. More precisely NESS is only supported on subspace spanned by states in the root(s) of wDAG.

While the NESS of XX model  has neither quantum nor classical correlation and describes trivial thermodynamics, the NESS of Fredkin model contains non-local information of paired spins. Since NESS counts correlated paired parenthesis, we expect $\rho$ having finite classical correlation which can be characterized by operator space entanglement entropy (OSEE). We do Schmidt decomposition of $\rho$ by expanding it in operator basis $\ket{i}\bra{j}_{A}$ of subregion A and $\ket{m}\bra{n}_{B}$ of B, i.e., $\rho = \sum_i p_i \rho^A_i \otimes \rho^B_i$ with normalization $\sum_i p_i^2 = 1$. The OSEE of $\rho$ is defined as $S_{\text{OP}}(\rho) = -\sum_i p_i^2 \log(p_i^2)$. By exact diagonalization for the Fredkin model with $N=8$ and $\beta=\log(3)$, we get the OSEE of $\rho$ given by $S_{\text{OP}}(\rho) \approx 0.599$, which is indeed a finite value, in contrast to $S_{\text{OP}}(\rho)=0$ for the XX model.

\section{Conclusions and outlook}
We unveiled that, by properly choosing jump operators, "finite temperature" Gibbs ensemble can emerge as the NESS for open quantum system described by Lindblad master equation. This construction is quite general for system with U(1) conservation laws. We explicitly construct NESS for models that possess one and two conserved charges, and find NESS can reach equilibrium even coupled with multiple bathes. An important future question is, how to construct similar Gibbs ensemble for system with non-Abelian symmetries like $\mathfrak{su}(2)$ of Heisenberg model or Temperley-Lieb model, a variant of pair-flip model with SU(3) symmetry. The difficulty lies in the construction of mappings between different irrep space. Besides, it is interesting to study time evolution under our construction, and examine to what extent quantum effect influence the dynamics, and how different placement or number of dissipation affect Liouvillian gap and the relaxation to the same NESS \citep{PhysRevE.92.042143, han2024exponentially}. Moreover, since we utilize Lindblad master equation which describes Markovian dynamics, it is also interesting how to apply our methods in non-Markovian cases, where there are some pioneer work \citep{PhysRevE.88.022121, 10.1063/1.3109898}. We will leave these questions to future work.

\begin{acknowledgments}
We especially thank C. G. Liang and B. Z. Zhou for insightful discussions and appreciate helpful communications with  X.L. Wang, Y. Zhang and K. Wang. The work is supported by National Key Research and Development Program of China (Grant No. 2023YFA1406704), the NSFC under Grants No. 12174436 and
No. T2121001 and the Strategic Priority Research Program of Chinese Academy of Sciences under Grant No. XDB33000000.
\end{acknowledgments}

\appendix
\section{More details for the Fredkin spin chain with boundary driving}

\begin{figure}[tb]
	\centering
	\includegraphics[width=1.\linewidth]{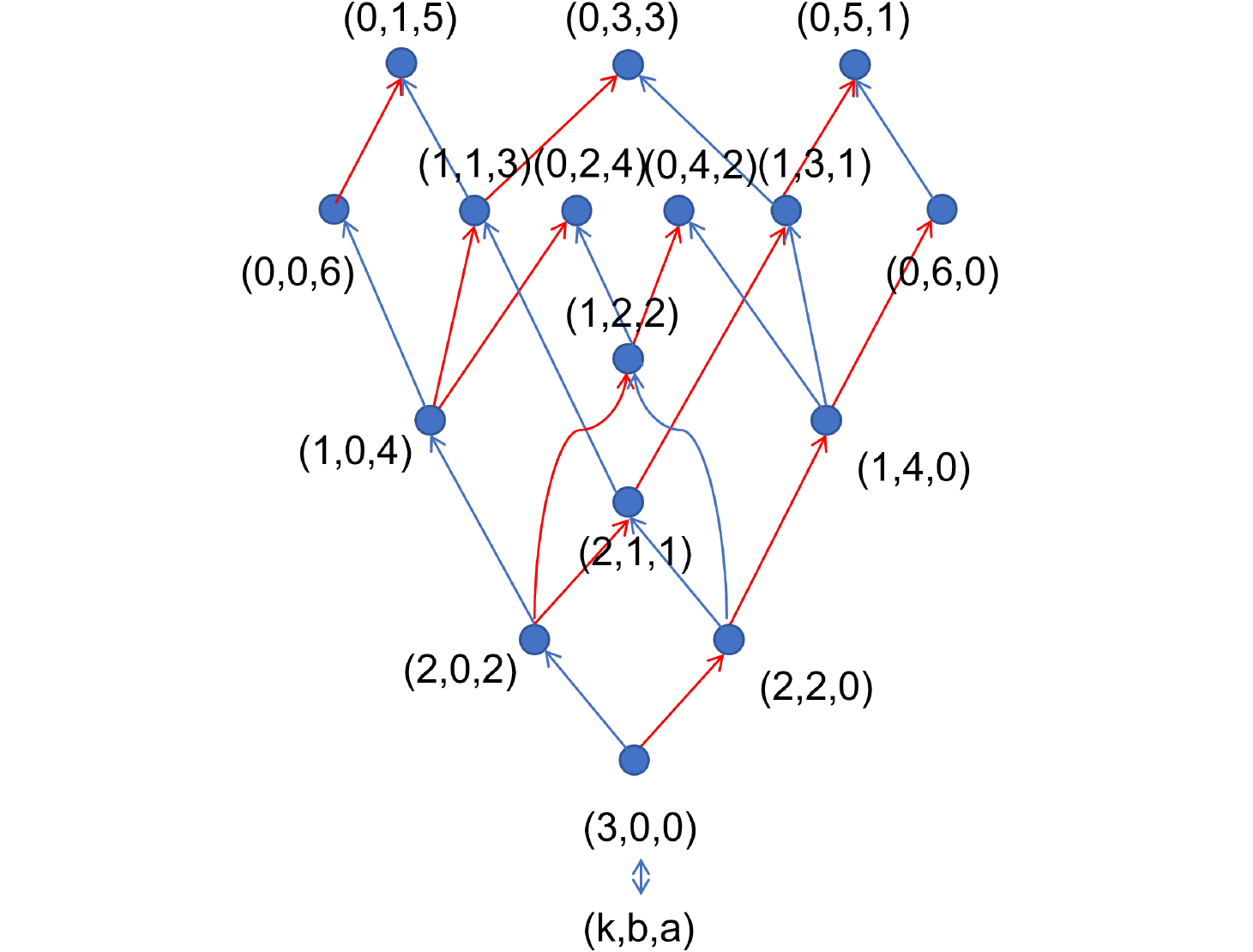}
	\caption{Illustration for NESS of Fredkin model with system length $N=6$. Tuples represent sector label by ($N^k$, $N^b$, $N^a$). Left blue arrow stands for $\sqrt{\gamma_2}L_1^{\dagger}$ and right red arrow stands for $\sqrt{\gamma_2}L_N$.}
	\label{fig:graph}
\end{figure}

The NESS of Fredkin model with different $N$ has similar hierarchical structure. To see it clearly, here we plot the hierarchical structure of NESS of Fredkin model for system with $N=6$ in Fig.{\ref{fig:graph}}, which is similar to Fig.1 in the main text.

Next we give an explanation of Eq.(\ref{eq:pairedp}) in the main text. In order to count paired parenthesis, we should make sure it is a pair of left and right parenthesis, so here we have $P^{\uparrow}_i P^{\downarrow}_{i+2m+1}$, which is separated by $2m$ distance.  Moreover, they cannot pair with any parenthesis inside. Note that if they are separated by odd distance, then one of them must be paired with one parenthesis between them. By projecting out unwanted configurations using projectors in product form, we arrive at Eq.(\ref{eq:pairedp}).

\begin{figure}[tb]
	\centering
	\includegraphics[width=1.0\linewidth]{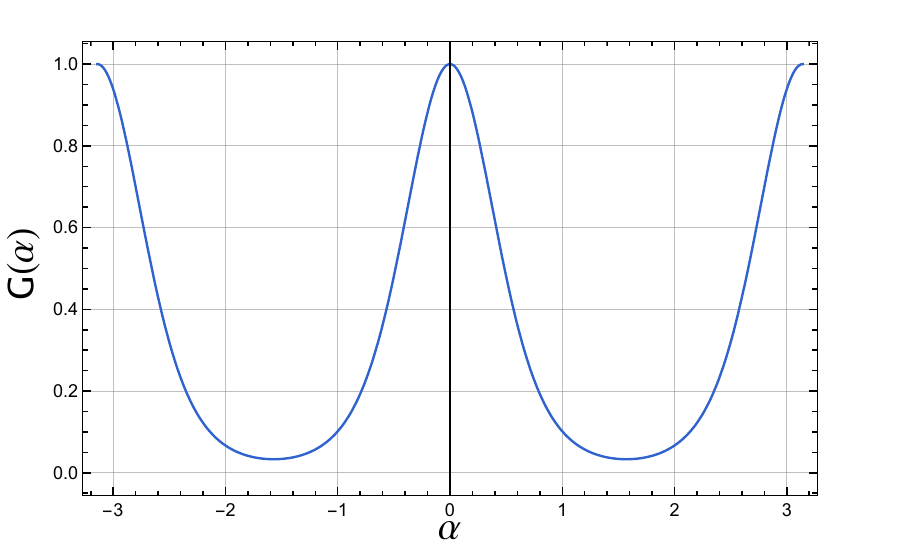}
	\caption{FCS of Fredkin model with N=200 and $\beta = \log(3)$.}
	\label{fig:fcsfredkin}
\end{figure}

Now we discuss the full counting statistics (FCS) of the Fredkin model with boundary driving. Since $[\tilde{H}, S^z]=0$, the FCS of operator $S^z=\sum_i \sigma^z_i$ is $\tilde{G}(\alpha) \propto \text{Tr}\lp e^{-\beta \tilde{H} + i \alpha S^z} \rp$ and the full counting function $G(\alpha) = \tilde{G}(\alpha)/\tilde{G}(0)$. From the form of $G$ we know $\text{Re}[G(\alpha)]=\text{Re}[G(-\alpha)]$ and $\text{Im}[G(\alpha)]=-\text{Im}[G(-\alpha)]$. Next we calculate the simple case $\text{Dis}=\{\sqrt{\gamma_1}L_1, \sqrt{\gamma_2}L_1^{\dagger},\sqrt{\gamma_2}L_N, \sqrt{\gamma_1}L_N^{\dagger} \}$ where $L_i = \sigma^{+}_i$. The size of a sector $\mathcal{K}(k,b,a)$ is \citep{doi:10.1142/S0129055X17500313fredkinchain} $
\text{D}_N (\mathcal{K}) = \left(\begin{array}{c}
	N \\
	N-k
\end{array}\right)-\left(\begin{array}{c}
	N \\
	N-k+1
\end{array}\right)
$
and all possible $(k,b,a)$ that satisfies $2k+b+a=N$ appear exactly once. The FCS is
\begin{widetext}
	\begin{equation}
		\begin{aligned}
			\tilde{G}(\alpha) &= \sum_{j=0}^{N/2} \sum_{m=0}^{j}  D_N\lp\frac{N}{2}-j, 2m, 2j-2m \rp e^{-\beta j+ i\alpha(4m - 2j) } + \\
			& \sum_{j=0}^{N/2-1} \sum_{m=0}^{j}  D_N\lp\frac{N}{2}-j-1, 2m+1, 2j-2m+1 \rp e^{-\beta (j+2) + i \alpha(4m - 2j) } \\
			&= \sum_{j=0}^{N/2} D_N\lp \frac{N}{2}-j\rp e^{-\beta j- 2i\alpha j } \frac{1-e^{4i\alpha (j+1)}}{1-e^{4i\alpha }} +
			\sum_{j=0}^{N/2-1} D_N\lp \frac{N}{2}-j-1\rp e^{-\beta (j+2)- 2i\alpha j } \frac{1-e^{4i\alpha (j+1)}}{1-e^{4i\alpha }} \\
			&=   \sum_{j=0}^{N/2}\frac{2j+1}{N+1} \left(\begin{array}{c}
				N+1 \\
				\frac{N}{2}-j
			\end{array}\right)  e^{-(\beta + 2i\alpha) j }\frac{1-e^{4i\alpha (j+1)}}{1-e^{4i\alpha }} + \sum_{j=0}^{N/2-1}  \frac{2j+3}{N+1} \left(\begin{array}{c}
				N+1 \\
				\frac{N}{2}-j-1
			\end{array}\right)  e^{-2\beta} e^{-(\beta + 2i\alpha) j } \frac{1-e^{4i\alpha (j+1)}}{1-e^{4i\alpha }} \\
			&=   \sum_{j=0}^{N/2}\frac{2j+1}{N+1} \left(\begin{array}{c}
				N+1 \\
				\frac{N}{2}-j
			\end{array}\right)  e^{-\beta j }\frac{\sin [ 2\alpha (j+1)]}{\sin (2\alpha)} + \sum_{j=0}^{N/2-1}  \frac{2j+3}{N+1} \left(\begin{array}{c}
				N+1 \\
				\frac{N}{2}-j-1
			\end{array}\right)  e^{-2\beta} e^{-\beta j }\frac{\sin [ 2\alpha (j+1)]}{\sin (2\alpha)}
		\end{aligned}
	\end{equation}
\end{widetext}
In the second equality we make use of definition of $D_N$ and sum over $m$. In the third equality we use property of binomial. This summation has no primary expression but can be efficiently calculated numerically. It can be checked that $G(\alpha)$ is real and can be represented as a composition of hypergeometric functions $_2F_1$. Direct numeric result is shown in Fig.\ref{fig:fcsfredkin}.
In the vicinity of $\alpha = 0$, $G(\alpha)\approx e^{-6\alpha^2}$, and in the vicinity of $\alpha = \frac{\pi}{2}$, $G(\alpha)\approx e^{-4\sin^2(\alpha)}$.

More generally when the set of jump operator $\text{Dis}=\{\sqrt{\gamma_1}L_1, \sqrt{\gamma_2}L_1^{\dagger},\sqrt{\gamma_3}L_N, \sqrt{\gamma_4}L_N^{\dagger} \}$ where $L_i = \sigma^{+}_i$, we still has $\rho\propto e^{-\beta \tilde{H}}$, but now it is ambiguous in defining $\beta$ because $\tilde{H}$ is no longer independent on $\beta$ we cannot define an unique temperature. For example,
\begin{equation}\label{eq:general_heff}
	\begin{aligned}
		\tilde{H} &= \frac{\beta_l}{\beta_l + \beta_r} N_a + \frac{\beta_r}{\beta_l + \beta_r} N_b  -\frac{1}{2}\lp \frac{3 \beta_l}{\beta_l + \beta_r}-1 \rp(-1)^{N^b} \\ &- \frac{1}{2}\lp \frac{3 \beta_r}{\beta_l + \beta_r}-1 \rp(-1)^{N^a} \, ,
	\end{aligned}
\end{equation}
where $\beta_l \equiv \log(\frac{\gamma_1}{\gamma_2})$, $\beta_r \equiv \log(\frac{\gamma_4}{\gamma_3})$ and $\beta = \frac{\beta_l+\beta_r}{2}$ (Here we use a different notation for $\beta_r$, which is differed by a minus sign in comparison with the definition in the main text.), or equivalently $T=\frac{2T_l T_r}{T_l + T_r}$. When $\beta_l = \beta_r$ (\ref{eq:general_heff}) is reduced to Eq.(11) differed by a constant factor in the main text. We can similarly calculate the FCS of magnetization
\begin{widetext}
	\begin{equation}
		\begin{aligned}
			\tilde{G}(\alpha) &= \sum_{j=0}^{N/2} \sum_{m=0}^{j}  D_N\lp\frac{N}{2}-j, 2m, 2j-2m \rp e^{-\beta_l (j-m) -\beta_r m+ i\alpha(4m - 2j) } + \\
			& \sum_{j=0}^{N/2-1} \sum_{m=0}^{j}  D_N\lp\frac{N}{2}-j-1, 2m+1, 2j-2m+1 \rp e^{-\beta_l (j-m+1) -\beta_r (m+1) + i \alpha(4m - 2j) } \, .
		\end{aligned}
	\end{equation}
\end{widetext}
But this time the imaginary part of $G$ is not zero anymore.

\section{Some other models}
Here we give more examples with the steady state characterized by a Gibbs ensemble  $\rho_{\text{NESS}} \propto e^{-\beta \tilde{H}}$.
\subsection{Spin-1 $t-J_z$}
We give a classical fragmentation example, which is spin-1 version of $t-J_z$ model and we only keep terms that are relevant to fragmentation (we drop $J_z$ terms). The model is described by
\begin{equation}
	\begin{aligned}
		H_{t-J_z} = &\sum_i \lp S^{+} (1-P^{\downarrow}) \rp_i \lp S^{-}P^{\uparrow} \rp_{i+1} \\&+ \lp S^{+} P^{\downarrow} \rp_i \lp S^{-}(1-P^{\uparrow}) \rp_{i+1} + h.c. \, .
	\end{aligned}
\end{equation}
We use $\ket{+}$, $\ket{-}$ and $\ket{0}$ to denote local onsite states, $P^{\downarrow} = \ket{-}\bra{-}$ and $P^{\uparrow} = \ket{+}\bra{+}$. This model is similar to a fermionic $t-J_z$ model without double occupation. An intuitive picture for this model is $\ket{+}, \ket{-}$ cannot pass through each other on the chain. Starting from an arbitrary computational basis, we can recognize its corresponding Krylov sector by neglecting all $\ket{0}$ and count the length of $N_{+}$ or $N_{-}$ sequence. All sectors can thus be labeled by a tuple $\lp N_{+}, N_{-}, N_{+}, ...\rp$ that describes number of $\ket{+}, \ket{-}$ in an alternating sequence of positive and negative integer, where $N_{+}$ is denoted by a  positive number and $N_{-}$ by a negative number. For example, when system length $N=4$, $(1,-1,1)$ labels the sector spanned by states $\ket{0,+,-,+}$, $ \ket{+,0,-,+}$, $\ket{+,-,0,+}$, $\ket{+,-,+,0}$.

\begin{figure}[tb]
	\centering
	\includegraphics[width=1.\linewidth]{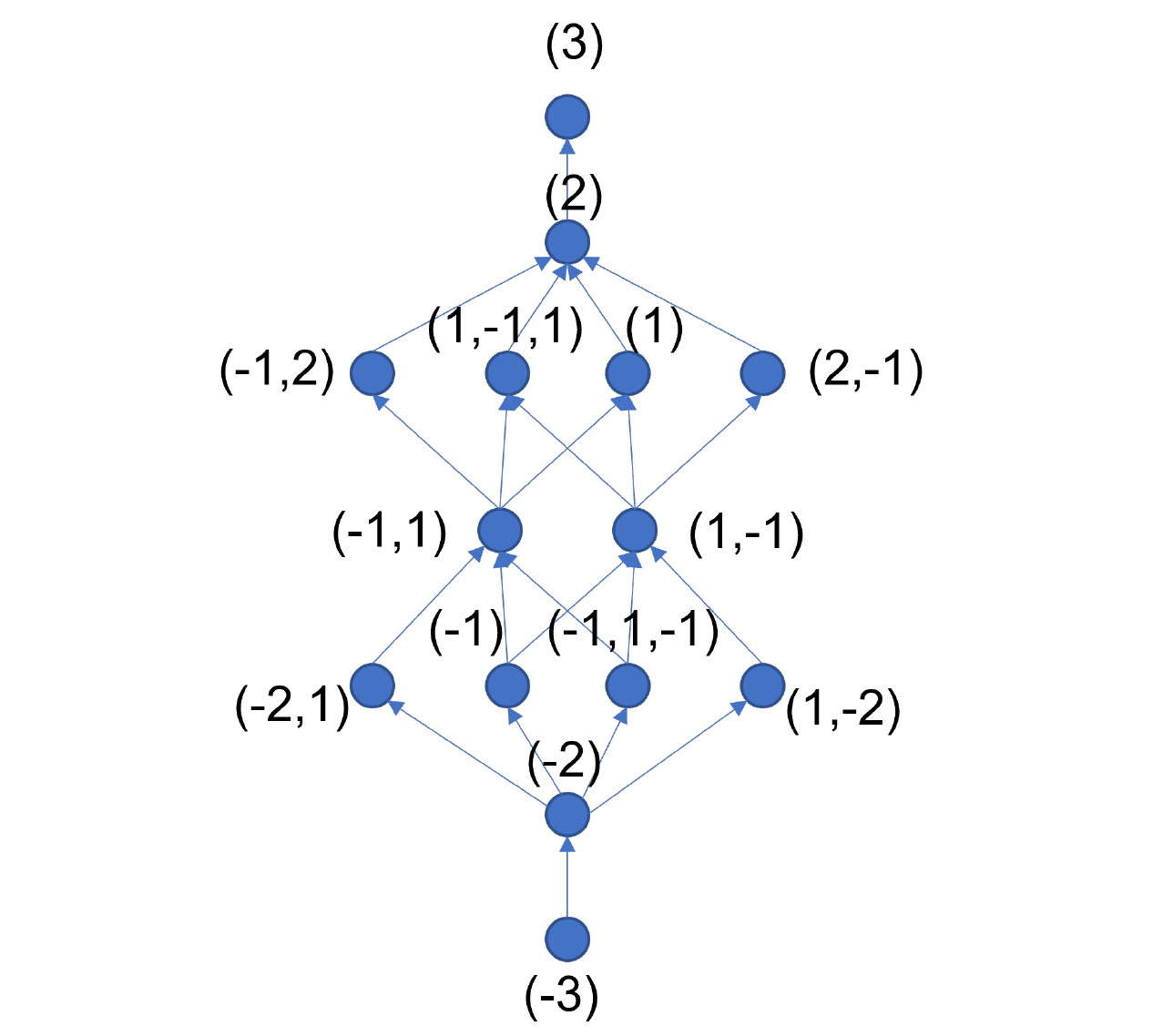}
	\caption{Structure of NESS of spin-1 $t-J_z$ model with $N=3$ under $L_\text{Dis2}$.}
	\label{fig:spin1graph3}
\end{figure}

Jump operators are not unique. For example, we can take simple ones like
\begin{equation}
	L_\text{Dis1} = \{L_{i,+}\equiv \sqrt{\gamma_1}S_{i}^{+} ,\, L_{i,-}\equiv \sqrt{\gamma_2} S_{i}^{-} \} \, ,
\end{equation}
acting on ends of the chain or the whole chain. More complicated we can also take
\begin{equation}
	\begin{aligned}
		L_\text{Dis2} = \{&L_{i,+}\equiv\frac{\sqrt{\gamma_1}}{2}\lp P^{\uparrow}_i S^{+}_{i+1}+ S^{+}_{i}P^{\uparrow}_{i+1} \rp , \\& L_{i,-}\equiv\frac{\sqrt{\gamma_2}}{2}\lp P^{\downarrow}_i S^{-}_{i+1}+ S^{-}_{i}P^{\downarrow}_{i+1} \rp,\, \\& \frac{\sqrt{\gamma_2}}{\sqrt{\gamma_1}}L_{i,+}^{\dagger},\, \frac{\sqrt{\gamma_1}}{\sqrt{\gamma_2}}L_{i,-}^{\dagger}\}
	\end{aligned}
\end{equation}
or
\begin{equation}
	\begin{aligned}
		L_\text{Dis3} = \{& L_{0,+} \equiv \sqrt{\gamma_1}S_1^{+}, L_{0,-} \equiv \sqrt{\gamma_2}S_1^{-}, \\ 
		& L_{i,+} \equiv \sqrt{\gamma_3}(S_i^+)^2(S_i^-)^2, \\&L_{i,-} \equiv \sqrt{\gamma_3}(S_i^-)^2(S_i^+)^2\}
	\end{aligned}
\end{equation}
In the second case jump operators behave as ladder operators in a non-unitary representation, that is, $L_{i,+}^{\dagger} \neq k L_{i,-}$. By this choice of jump operators we exclude state $\ket{0,0,0,...}$ because it is a frozen state of both Hamiltonian and jump operators. The remaining all states form NESS and is shown in Fig.\ref{fig:spin1graph3} for $N=3$.

Then $\rho_{\text{NESS}}\propto e^{-\beta \tilde{H}}$ with $\beta = \log(\frac{\gamma_1}{\gamma_2}) $ and the effective Hamiltonian is just the total magnetization
\begin{equation}
	\tilde{H} = -\sum_{i} S^z_i \equiv -S^z_{\text{tot}} \, .
\end{equation}

\subsection{Spin-1 pair-flip}
The Hamiltonian of spin-1 pair-flip model reads \citep{moudgalya2022hilbert}
\begin{equation}
	H_{\text{PF}} = \sum_{i} \sum_{\alpha, \beta\in \{+,0,-\}} g_{i,i+1}^{\alpha \beta} \lp \ket{\alpha\alpha}\bra{\beta\beta} \rp_{i,i+1} + h.c. \, ,
\end{equation}
where $g_{i,i+1}^{\alpha \beta}$ is random and we only keep terms relevant to fragmentation. Each Krylov sector is labeled by its dot pattern. From an arbitrary state, we can derive its dot pattern by consecutively link two adjacent spin with identical orientation and drop them. The final remaining spins form the dot pattern. For example, (a) $\ket{+--+}\rightarrow \ket{++} \rightarrow \ket{}$, an empty pattern; (b) $\ket{0-00+}\rightarrow \ket{0-+}$.

Jump operators are
\begin{equation}
		\begin{aligned}
		L_\text{Dis} = \{ & L_{i,+}\equiv\sqrt{\gamma_1}S_{2i+1}^{+} ,\, L_{i,-}\equiv\sqrt{\gamma_1} S_{2i+2}^{-},\, \\ & \frac{\sqrt{\gamma_2}}{\sqrt{\gamma_1}}L_{i,+}^{\dagger},\, \frac{\sqrt{\gamma_2}}{\sqrt{\gamma_1}}L_{i,-}^{\dagger} \} \, ,
	\end{aligned}
\end{equation}

where by definition $L_{i,+}$ acts on any odd sites and $L_{i,-}$ acts on any even sites. This choice of jump operators is based on $U(1)$ charges of pair-flip model, $Q_a = \sum_i (-1)^i \ket{a}\bra{a}_i$.
\begin{figure}[tb]
	\centering
	\includegraphics[width=1.\linewidth]{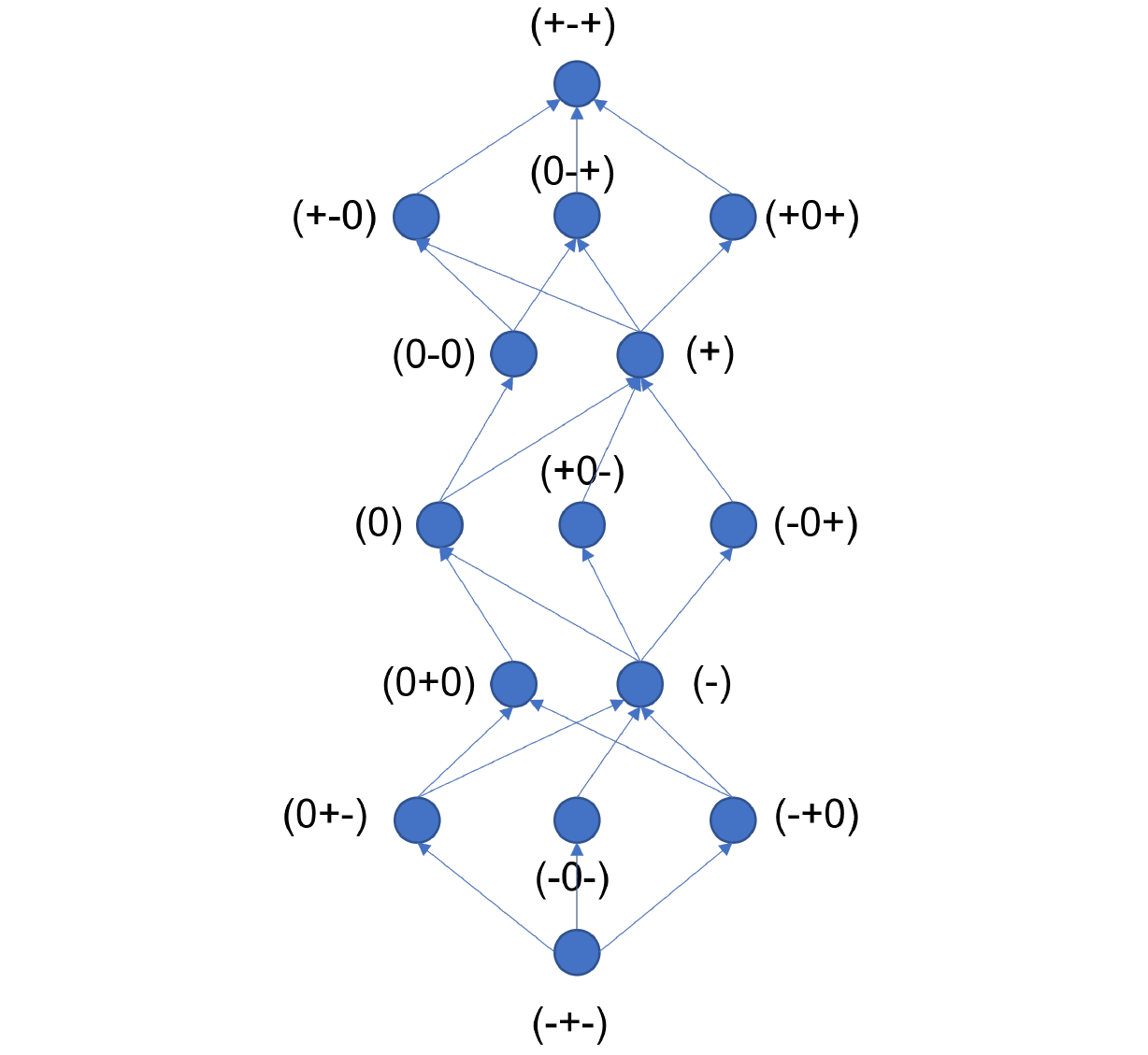}
	\caption{Structure of NESS of spin-1 pair-flip model with N=3.}
	\label{fig:spin1pf3}
\end{figure}

Then $\rho_{\text{NESS}}\propto e^{-\beta \tilde{H}}$ with $\beta = \log(\frac{\gamma_1}{\gamma_2}) $ and the effective Hamiltonian is just the staggered magnetization
\begin{equation}
	\tilde{H} = \sum_{i} \lp -1\rp^{i} S^z_i \equiv S^z_{\text{st}} \, .
\end{equation}
An illustration for the structure of NESS is shown in Fig.\ref{fig:spin1pf3}.

\subsection{Spin-1 dipole conserving}
The Hamiltonian of spin-1 dipole conserving model \citep{PhysRevX.10.011047dipole} reads
\begin{equation}
	H_{\text{DC}} = \sum_{i} S_i^- \lp S_{i+1}^{+}\rp^2 S_{i+2}^- + h.c. \, .
\end{equation}
This model has two $U(1)$ charges: total magnetization and total dipole. In this case different charge can couple to bath with different "temperature", similar to Fredkin model in the main text. Jump operators are
\begin{equation}
	\begin{aligned}
		L_\text{Dis} = \{&L_{0,+} \equiv \sqrt{\gamma_1}S_1^{+}, L_{0,-} \equiv \sqrt{\gamma_2}S_1^{-}, \\ 
		&L_{i,+} \equiv \sqrt{\gamma_3}S_i^- S_{i+1}^+ P_{i,i+1}, \\ &L_{i,+} \equiv \sqrt{\gamma_4} S_i^+ S_{i+1}^- P_{i,i+1} \}
	\end{aligned}
\end{equation}
where $P_{i,i+1}$ is projector into subspace spanned by $\ket{+0}$, $\ket{-0}$, $\ket{0+}$ and $\ket{0-}$. In simple case where $\frac{\gamma_3}{\gamma_4}=\frac{\gamma_1}{\gamma_2}$, $\rho_{\text{NESS}}\propto e^{-\beta \tilde{H}}$ with $\beta = \log(\frac{\gamma_3}{\gamma_4}) $ and the effective Hamiltonian is the total dipole
\begin{equation}
	\tilde{H} = -\sum_{j} j S^z_j  \, .
\end{equation}
In general case $\beta_1 = \log(\frac{\gamma_1}{\gamma_2}) $ and $\beta_2 = \log(\frac{\gamma_3}{\gamma_4}) $, the effective Hamiltonian is proportional to a linear composition of total magnetization and total dipole
\begin{equation}
	\tilde{H}\propto -\sum_{j} \lp \beta_2 \, j S^z_j + (\beta_1-\beta_2)S^z_j \rp\, .
\end{equation}

\bibliography{lib}

\begin{thebibliography}{55}%
\makeatletter
\providecommand \@ifxundefined [1]{%
 \@ifx{#1\undefined}
}%
\providecommand \@ifnum [1]{%
 \ifnum #1\expandafter \@firstoftwo
 \else \expandafter \@secondoftwo
 \fi
}%
\providecommand \@ifx [1]{%
 \ifx #1\expandafter \@firstoftwo
 \else \expandafter \@secondoftwo
 \fi
}%
\providecommand \natexlab [1]{#1}%
\providecommand \enquote  [1]{``#1''}%
\providecommand \bibnamefont  [1]{#1}%
\providecommand \bibfnamefont [1]{#1}%
\providecommand \citenamefont [1]{#1}%
\providecommand \href@noop [0]{\@secondoftwo}%
\providecommand \href [0]{\begingroup \@sanitize@url \@href}%
\providecommand \@href[1]{\@@startlink{#1}\@@href}%
\providecommand \@@href[1]{\endgroup#1\@@endlink}%
\providecommand \@sanitize@url [0]{\catcode `\\12\catcode `\$12\catcode
  `\&12\catcode `\#12\catcode `\^12\catcode `\_12\catcode `\%12\relax}%
\providecommand \@@startlink[1]{}%
\providecommand \@@endlink[0]{}%
\providecommand \url  [0]{\begingroup\@sanitize@url \@url }%
\providecommand \@url [1]{\endgroup\@href {#1}{\urlprefix }}%
\providecommand \urlprefix  [0]{URL }%
\providecommand \Eprint [0]{\href }%
\providecommand \doibase [0]{https://doi.org/}%
\providecommand \selectlanguage [0]{\@gobble}%
\providecommand \bibinfo  [0]{\@secondoftwo}%
\providecommand \bibfield  [0]{\@secondoftwo}%
\providecommand \translation [1]{[#1]}%
\providecommand \BibitemOpen [0]{}%
\providecommand \bibitemStop [0]{}%
\providecommand \bibitemNoStop [0]{.\EOS\space}%
\providecommand \EOS [0]{\spacefactor3000\relax}%
\providecommand \BibitemShut  [1]{\csname bibitem#1\endcsname}%
\let\auto@bib@innerbib\@empty
\bibitem [{\citenamefont {Gorini}\ \emph {et~al.}(1976)\citenamefont {Gorini},
  \citenamefont {Kossakowski},\ and\ \citenamefont
  {Sudarshan}}]{10.1063/1.522979GKS}%
  \BibitemOpen
  \bibfield  {author} {\bibinfo {author} {\bibfnamefont {V.}~\bibnamefont
  {Gorini}}, \bibinfo {author} {\bibfnamefont {A.}~\bibnamefont
  {Kossakowski}},\ and\ \bibinfo {author} {\bibfnamefont {E.~C.~G.}\
  \bibnamefont {Sudarshan}},\ }\bibfield  {title} {\bibinfo {title}
  {{Completely positive dynamical semigroups of N-level systems}},\ }\href
  {https://doi.org/10.1063/1.522979} {\bibfield  {journal} {\bibinfo  {journal}
  {Journal of Mathematical Physics}\ }\textbf {\bibinfo {volume} {17}},\
  \bibinfo {pages} {821} (\bibinfo {year} {1976})}\BibitemShut {NoStop}%
\bibitem [{\citenamefont {Lindblad}(1976)}]{Lindblad1976}%
  \BibitemOpen
  \bibfield  {author} {\bibinfo {author} {\bibfnamefont {G.}~\bibnamefont
  {Lindblad}},\ }\bibfield  {title} {\bibinfo {title} {On the generators of
  quantum dynamical semigroups},\ }\href {https://doi.org/10.1007/BF01608499}
  {\bibfield  {journal} {\bibinfo  {journal} {Communications in Mathematical
  Physics}\ }\textbf {\bibinfo {volume} {48}},\ \bibinfo {pages} {119}
  (\bibinfo {year} {1976})}\BibitemShut {NoStop}%
\bibitem [{\citenamefont {Breuer}\ and\ \citenamefont
  {Petruccione}(2007)}]{oqsbook}%
  \BibitemOpen
  \bibfield  {author} {\bibinfo {author} {\bibfnamefont {H.-P.}\ \bibnamefont
  {Breuer}}\ and\ \bibinfo {author} {\bibfnamefont {F.}~\bibnamefont
  {Petruccione}},\ }\href
  {https://doi.org/10.1093/acprof:oso/9780199213900.001.0001} {\emph {\bibinfo
  {title} {{The Theory of Open Quantum Systems}}}}\ (\bibinfo  {publisher}
  {Oxford University Press},\ \bibinfo {year} {2007})\BibitemShut {NoStop}%
\bibitem [{\citenamefont {Kamenev}(2011)}]{Kamenev_2011}%
  \BibitemOpen
  \bibfield  {author} {\bibinfo {author} {\bibfnamefont {A.}~\bibnamefont
  {Kamenev}},\ }\href@noop {} {\emph {\bibinfo {title} {Field Theory of
  Non-Equilibrium Systems}}}\ (\bibinfo  {publisher} {Cambridge University
  Press},\ \bibinfo {year} {2011})\BibitemShut {NoStop}%
\bibitem [{\citenamefont {Derrida}(2007)}]{Derrida_2007}%
  \BibitemOpen
  \bibfield  {author} {\bibinfo {author} {\bibfnamefont {B.}~\bibnamefont
  {Derrida}},\ }\bibfield  {title} {\bibinfo {title} {Non-equilibrium steady
  states: fluctuations and large deviations of the density and of the
  current},\ }\href {https://doi.org/10.1088/1742-5468/2007/07/P07023}
  {\bibfield  {journal} {\bibinfo  {journal} {Journal of Statistical Mechanics:
  Theory and Experiment}\ }\textbf {\bibinfo {volume} {2007}},\ \bibinfo
  {pages} {P07023} (\bibinfo {year} {2007})}\BibitemShut {NoStop}%
\bibitem [{\citenamefont {Garrahan}(2018)}]{GARRAHAN2018130}%
  \BibitemOpen
  \bibfield  {author} {\bibinfo {author} {\bibfnamefont {J.~P.}\ \bibnamefont
  {Garrahan}},\ }\bibfield  {title} {\bibinfo {title} {Aspects of
  non-equilibrium in classical and quantum systems: Slow relaxation and
  glasses, dynamical large deviations, quantum non-ergodicity, and open quantum
  dynamics},\ }\href
  {https://doi.org/https://doi.org/10.1016/j.physa.2017.12.149} {\bibfield
  {journal} {\bibinfo  {journal} {Physica A: Statistical Mechanics and its
  Applications}\ }\textbf {\bibinfo {volume} {504}},\ \bibinfo {pages} {130}
  (\bibinfo {year} {2018})},\ \bibinfo {note} {lecture Notes of the 14th
  International Summer School on Fundamental Problems in Statistical
  Physics}\BibitemShut {NoStop}%
\bibitem [{\citenamefont {Prosen}(2008)}]{Prosen_2008}%
  \BibitemOpen
  \bibfield  {author} {\bibinfo {author} {\bibfnamefont {T.}~\bibnamefont
  {Prosen}},\ }\bibfield  {title} {\bibinfo {title} {Third quantization: a
  general method to solve master equations for quadratic open fermi systems},\
  }\href {https://doi.org/10.1088/1367-2630/10/4/043026} {\bibfield  {journal}
  {\bibinfo  {journal} {New Journal of Physics}\ }\textbf {\bibinfo {volume}
  {10}},\ \bibinfo {pages} {043026} (\bibinfo {year} {2008})}\BibitemShut
  {NoStop}%
\bibitem [{\citenamefont {Verstraete}\ \emph {et~al.}(2009)\citenamefont
  {Verstraete}, \citenamefont {Wolf},\ and\ \citenamefont
  {Cirac}}]{Verstraete2009}%
  \BibitemOpen
  \bibfield  {author} {\bibinfo {author} {\bibfnamefont {F.}~\bibnamefont
  {Verstraete}}, \bibinfo {author} {\bibfnamefont {M.~M.}\ \bibnamefont
  {Wolf}},\ and\ \bibinfo {author} {\bibfnamefont {J.~I.}\ \bibnamefont
  {Cirac}},\ }\bibfield  {title} {\bibinfo {title} {Quantum computation and
  quantum-state engineering driven by dissipation},\ }\href
  {https://doi.org/10.1038/nphys1342} {\bibfield  {journal} {\bibinfo
  {journal} {Nature Physics}\ }\textbf {\bibinfo {volume} {5}},\ \bibinfo
  {pages} {633} (\bibinfo {year} {2009})}\BibitemShut {NoStop}%
\bibitem [{\citenamefont {Dubois}\ \emph {et~al.}(2023)\citenamefont {Dubois},
  \citenamefont {Saalmann},\ and\ \citenamefont
  {Rost}}]{PhysRevResearch.5.L012003}%
  \BibitemOpen
  \bibfield  {author} {\bibinfo {author} {\bibfnamefont {J.}~\bibnamefont
  {Dubois}}, \bibinfo {author} {\bibfnamefont {U.}~\bibnamefont {Saalmann}},\
  and\ \bibinfo {author} {\bibfnamefont {J.~M.}\ \bibnamefont {Rost}},\
  }\bibfield  {title} {\bibinfo {title} {Symmetry-induced decoherence-free
  subspaces},\ }\href {https://doi.org/10.1103/PhysRevResearch.5.L012003}
  {\bibfield  {journal} {\bibinfo  {journal} {Phys. Rev. Res.}\ }\textbf
  {\bibinfo {volume} {5}},\ \bibinfo {pages} {L012003} (\bibinfo {year}
  {2023})}\BibitemShut {NoStop}%
\bibitem [{\citenamefont {Liu}\ \emph {et~al.}(2024)\citenamefont {Liu},
  \citenamefont {Wang}, \citenamefont {Yang}, \citenamefont {Jie},\ and\
  \citenamefont {Wang}}]{PhysRevLett.132.216301}%
  \BibitemOpen
  \bibfield  {author} {\bibinfo {author} {\bibfnamefont {Y.}~\bibnamefont
  {Liu}}, \bibinfo {author} {\bibfnamefont {Z.}~\bibnamefont {Wang}}, \bibinfo
  {author} {\bibfnamefont {C.}~\bibnamefont {Yang}}, \bibinfo {author}
  {\bibfnamefont {J.}~\bibnamefont {Jie}},\ and\ \bibinfo {author}
  {\bibfnamefont {Y.}~\bibnamefont {Wang}},\ }\bibfield  {title} {\bibinfo
  {title} {Dissipation-induced extended-localized transition},\ }\href
  {https://doi.org/10.1103/PhysRevLett.132.216301} {\bibfield  {journal}
  {\bibinfo  {journal} {Phys. Rev. Lett.}\ }\textbf {\bibinfo {volume} {132}},\
  \bibinfo {pages} {216301} (\bibinfo {year} {2024})}\BibitemShut {NoStop}%
\bibitem [{\citenamefont {Bu{\v{c}}a}\ \emph {et~al.}(2019)\citenamefont
  {Bu{\v{c}}a}, \citenamefont {Tindall},\ and\ \citenamefont
  {Jaksch}}]{buvca2019non}%
  \BibitemOpen
  \bibfield  {author} {\bibinfo {author} {\bibfnamefont {B.}~\bibnamefont
  {Bu{\v{c}}a}}, \bibinfo {author} {\bibfnamefont {J.}~\bibnamefont
  {Tindall}},\ and\ \bibinfo {author} {\bibfnamefont {D.}~\bibnamefont
  {Jaksch}},\ }\bibfield  {title} {\bibinfo {title} {Non-stationary coherent
  quantum many-body dynamics through dissipation},\ }\href@noop {} {\bibfield
  {journal} {\bibinfo  {journal} {Nature Communications}\ }\textbf {\bibinfo
  {volume} {10}},\ \bibinfo {pages} {1730} (\bibinfo {year}
  {2019})}\BibitemShut {NoStop}%
\bibitem [{\citenamefont {Rakovszky}\ \emph {et~al.}(2023)\citenamefont
  {Rakovszky}, \citenamefont {Gopalakrishnan},\ and\ \citenamefont {von
  Keyserlingk}}]{rakovszky2023defining}%
  \BibitemOpen
  \bibfield  {author} {\bibinfo {author} {\bibfnamefont {T.}~\bibnamefont
  {Rakovszky}}, \bibinfo {author} {\bibfnamefont {S.}~\bibnamefont
  {Gopalakrishnan}},\ and\ \bibinfo {author} {\bibfnamefont {C.}~\bibnamefont
  {von Keyserlingk}},\ }\href@noop {} {\bibinfo {title} {Defining stable phases
  of open quantum systems}} (\bibinfo {year} {2023}),\ \Eprint
  {https://arxiv.org/abs/2308.15495} {arXiv:2308.15495 [quant-ph]} \BibitemShut
  {NoStop}%
\bibitem [{\citenamefont {Wang}\ \emph {et~al.}(2023)\citenamefont {Wang},
  \citenamefont {Fang},\ and\ \citenamefont {Ren}}]{wang2023superdiffusive}%
  \BibitemOpen
  \bibfield  {author} {\bibinfo {author} {\bibfnamefont {Y.-P.}\ \bibnamefont
  {Wang}}, \bibinfo {author} {\bibfnamefont {C.}~\bibnamefont {Fang}},\ and\
  \bibinfo {author} {\bibfnamefont {J.}~\bibnamefont {Ren}},\ }\href@noop {}
  {\bibinfo {title} {Superdiffusive transport in quasi-particle dephasing
  models}} (\bibinfo {year} {2023}),\ \Eprint
  {https://arxiv.org/abs/2310.03069} {arXiv:2310.03069 [cond-mat.stat-mech]}
  \BibitemShut {NoStop}%
\bibitem [{\citenamefont {Haga}\ \emph {et~al.}(2021)\citenamefont {Haga},
  \citenamefont {Nakagawa}, \citenamefont {Hamazaki},\ and\ \citenamefont
  {Ueda}}]{PhysRevLett.127.070402}%
  \BibitemOpen
  \bibfield  {author} {\bibinfo {author} {\bibfnamefont {T.}~\bibnamefont
  {Haga}}, \bibinfo {author} {\bibfnamefont {M.}~\bibnamefont {Nakagawa}},
  \bibinfo {author} {\bibfnamefont {R.}~\bibnamefont {Hamazaki}},\ and\
  \bibinfo {author} {\bibfnamefont {M.}~\bibnamefont {Ueda}},\ }\bibfield
  {title} {\bibinfo {title} {Liouvillian skin effect: Slowing down of
  relaxation processes without gap closing},\ }\href
  {https://doi.org/10.1103/PhysRevLett.127.070402} {\bibfield  {journal}
  {\bibinfo  {journal} {Phys. Rev. Lett.}\ }\textbf {\bibinfo {volume} {127}},\
  \bibinfo {pages} {070402} (\bibinfo {year} {2021})}\BibitemShut {NoStop}%
\bibitem [{\citenamefont {Schmolke}\ and\ \citenamefont
  {Lutz}(2022)}]{PhysRevLett.129.250601}%
  \BibitemOpen
  \bibfield  {author} {\bibinfo {author} {\bibfnamefont {F.}~\bibnamefont
  {Schmolke}}\ and\ \bibinfo {author} {\bibfnamefont {E.}~\bibnamefont
  {Lutz}},\ }\bibfield  {title} {\bibinfo {title} {Noise-induced quantum
  synchronization},\ }\href {https://doi.org/10.1103/PhysRevLett.129.250601}
  {\bibfield  {journal} {\bibinfo  {journal} {Phys. Rev. Lett.}\ }\textbf
  {\bibinfo {volume} {129}},\ \bibinfo {pages} {250601} (\bibinfo {year}
  {2022})}\BibitemShut {NoStop}%
\bibitem [{\citenamefont {Zhou}\ \emph {et~al.}(2022)\citenamefont {Zhou},
  \citenamefont {Wang},\ and\ \citenamefont {Chen}}]{PhysRevB.106.064203}%
  \BibitemOpen
  \bibfield  {author} {\bibinfo {author} {\bibfnamefont {B.}~\bibnamefont
  {Zhou}}, \bibinfo {author} {\bibfnamefont {X.}~\bibnamefont {Wang}},\ and\
  \bibinfo {author} {\bibfnamefont {S.}~\bibnamefont {Chen}},\ }\bibfield
  {title} {\bibinfo {title} {Exponential size scaling of the liouvillian gap in
  boundary-dissipated systems with anderson localization},\ }\href
  {https://doi.org/10.1103/PhysRevB.106.064203} {\bibfield  {journal} {\bibinfo
   {journal} {Phys. Rev. B}\ }\textbf {\bibinfo {volume} {106}},\ \bibinfo
  {pages} {064203} (\bibinfo {year} {2022})}\BibitemShut {NoStop}%
\bibitem [{\citenamefont {Zheng}\ \emph {et~al.}(2023)\citenamefont {Zheng},
  \citenamefont {Wang},\ and\ \citenamefont {Chen}}]{PhysRevB.108.024404}%
  \BibitemOpen
  \bibfield  {author} {\bibinfo {author} {\bibfnamefont {Z.-Y.}\ \bibnamefont
  {Zheng}}, \bibinfo {author} {\bibfnamefont {X.}~\bibnamefont {Wang}},\ and\
  \bibinfo {author} {\bibfnamefont {S.}~\bibnamefont {Chen}},\ }\bibfield
  {title} {\bibinfo {title} {Exact solution of the boundary-dissipated
  transverse field ising model: Structure of the liouvillian spectrum and
  dynamical duality},\ }\href {https://doi.org/10.1103/PhysRevB.108.024404}
  {\bibfield  {journal} {\bibinfo  {journal} {Phys. Rev. B}\ }\textbf {\bibinfo
  {volume} {108}},\ \bibinfo {pages} {024404} (\bibinfo {year}
  {2023})}\BibitemShut {NoStop}%
\bibitem [{\citenamefont {Rall}\ \emph {et~al.}(2023)\citenamefont {Rall},
  \citenamefont {Wang},\ and\ \citenamefont {Wocjan}}]{Rall2023thermalstate}%
  \BibitemOpen
  \bibfield  {author} {\bibinfo {author} {\bibfnamefont {P.}~\bibnamefont
  {Rall}}, \bibinfo {author} {\bibfnamefont {C.}~\bibnamefont {Wang}},\ and\
  \bibinfo {author} {\bibfnamefont {P.}~\bibnamefont {Wocjan}},\ }\bibfield
  {title} {\bibinfo {title} {Thermal {S}tate {P}reparation via {R}ounding
  {P}romises},\ }\href {https://doi.org/10.22331/q-2023-10-10-1132} {\bibfield
  {journal} {\bibinfo  {journal} {{Quantum}}\ }\textbf {\bibinfo {volume}
  {7}},\ \bibinfo {pages} {1132} (\bibinfo {year} {2023})}\BibitemShut
  {NoStop}%
\bibitem [{\citenamefont {Chen}\ \emph {et~al.}(2023)\citenamefont {Chen},
  \citenamefont {Kastoryano}, \citenamefont {Brandão},\ and\ \citenamefont
  {Gilyén}}]{chen2023quantum}%
  \BibitemOpen
  \bibfield  {author} {\bibinfo {author} {\bibfnamefont {C.-F.}\ \bibnamefont
  {Chen}}, \bibinfo {author} {\bibfnamefont {M.~J.}\ \bibnamefont
  {Kastoryano}}, \bibinfo {author} {\bibfnamefont {F.~G. S.~L.}\ \bibnamefont
  {Brandão}},\ and\ \bibinfo {author} {\bibfnamefont {A.}~\bibnamefont
  {Gilyén}},\ }\href@noop {} {\bibinfo {title} {Quantum thermal state
  preparation}} (\bibinfo {year} {2023}),\ \Eprint
  {https://arxiv.org/abs/2303.18224} {arXiv:2303.18224 [quant-ph]} \BibitemShut
  {NoStop}%
\bibitem [{\citenamefont {Ding}\ \emph {et~al.}(2024)\citenamefont {Ding},
  \citenamefont {Li},\ and\ \citenamefont {Lin}}]{ding2024efficient}%
  \BibitemOpen
  \bibfield  {author} {\bibinfo {author} {\bibfnamefont {Z.}~\bibnamefont
  {Ding}}, \bibinfo {author} {\bibfnamefont {B.}~\bibnamefont {Li}},\ and\
  \bibinfo {author} {\bibfnamefont {L.}~\bibnamefont {Lin}},\ }\href@noop {}
  {\bibinfo {title} {Efficient quantum gibbs samplers with
  kubo--martin--schwinger detailed balance condition}} (\bibinfo {year}
  {2024}),\ \Eprint {https://arxiv.org/abs/2404.05998} {arXiv:2404.05998
  [quant-ph]} \BibitemShut {NoStop}%
\bibitem [{\citenamefont {Guo}\ \emph {et~al.}(2024)\citenamefont {Guo},
  \citenamefont {Hart}, \citenamefont {Chen}, \citenamefont {Friedman},\ and\
  \citenamefont {Lucas}}]{guo2024designing}%
  \BibitemOpen
  \bibfield  {author} {\bibinfo {author} {\bibfnamefont {J.}~\bibnamefont
  {Guo}}, \bibinfo {author} {\bibfnamefont {O.}~\bibnamefont {Hart}}, \bibinfo
  {author} {\bibfnamefont {C.-F.}\ \bibnamefont {Chen}}, \bibinfo {author}
  {\bibfnamefont {A.~J.}\ \bibnamefont {Friedman}},\ and\ \bibinfo {author}
  {\bibfnamefont {A.}~\bibnamefont {Lucas}},\ }\href@noop {} {\bibinfo {title}
  {Designing open quantum systems with known steady states: Davies generators
  and beyond}} (\bibinfo {year} {2024}),\ \Eprint
  {https://arxiv.org/abs/2404.14538} {arXiv:2404.14538 [quant-ph]} \BibitemShut
  {NoStop}%
\bibitem [{\citenamefont {Moudgalya}\ and\ \citenamefont
  {Motrunich}(2022)}]{moudgalya2022hilbert}%
  \BibitemOpen
  \bibfield  {author} {\bibinfo {author} {\bibfnamefont {S.}~\bibnamefont
  {Moudgalya}}\ and\ \bibinfo {author} {\bibfnamefont {O.~I.}\ \bibnamefont
  {Motrunich}},\ }\bibfield  {title} {\bibinfo {title} {Hilbert space
  fragmentation and commutant algebras},\ }\href@noop {} {\bibfield  {journal}
  {\bibinfo  {journal} {Physical Review X}\ }\textbf {\bibinfo {volume} {12}},\
  \bibinfo {pages} {011050} (\bibinfo {year} {2022})}\BibitemShut {NoStop}%
\bibitem [{\citenamefont {Haag}\ \emph {et~al.}(1967)\citenamefont {Haag},
  \citenamefont {Hugenholtz},\ and\ \citenamefont {Winnink}}]{Haag1967}%
  \BibitemOpen
  \bibfield  {author} {\bibinfo {author} {\bibfnamefont {R.}~\bibnamefont
  {Haag}}, \bibinfo {author} {\bibfnamefont {N.~M.}\ \bibnamefont
  {Hugenholtz}},\ and\ \bibinfo {author} {\bibfnamefont {M.}~\bibnamefont
  {Winnink}},\ }\bibfield  {title} {\bibinfo {title} {On the equilibrium states
  in quantum statistical mechanics},\ }\href
  {https://doi.org/10.1007/BF01646342} {\bibfield  {journal} {\bibinfo
  {journal} {Communications in Mathematical Physics}\ }\textbf {\bibinfo
  {volume} {5}},\ \bibinfo {pages} {215} (\bibinfo {year} {1967})}\BibitemShut
  {NoStop}%
\bibitem [{\citenamefont {Firanko}\ \emph {et~al.}(2024)\citenamefont
  {Firanko}, \citenamefont {Goldstein},\ and\ \citenamefont
  {Arad}}]{10.1063/5.0167353}%
  \BibitemOpen
  \bibfield  {author} {\bibinfo {author} {\bibfnamefont {R.}~\bibnamefont
  {Firanko}}, \bibinfo {author} {\bibfnamefont {M.}~\bibnamefont {Goldstein}},\
  and\ \bibinfo {author} {\bibfnamefont {I.}~\bibnamefont {Arad}},\ }\bibfield
  {title} {\bibinfo {title} {{Area law for steady states of detailed-balance
  local Lindbladians}},\ }\href {https://doi.org/10.1063/5.0167353} {\bibfield
  {journal} {\bibinfo  {journal} {Journal of Mathematical Physics}\ }\textbf
  {\bibinfo {volume} {65}},\ \bibinfo {pages} {051901} (\bibinfo {year}
  {2024})}\BibitemShut {NoStop}%
\bibitem [{\citenamefont {Bardet}\ \emph {et~al.}(2023)\citenamefont {Bardet},
  \citenamefont {Capel}, \citenamefont {Gao}, \citenamefont {Lucia},
  \citenamefont {P\'erez-Garc\'{\i}a},\ and\ \citenamefont
  {Rouz\'e}}]{PhysRevLett.130.060401}%
  \BibitemOpen
  \bibfield  {author} {\bibinfo {author} {\bibfnamefont {I.}~\bibnamefont
  {Bardet}}, \bibinfo {author} {\bibfnamefont {A.}~\bibnamefont {Capel}},
  \bibinfo {author} {\bibfnamefont {L.}~\bibnamefont {Gao}}, \bibinfo {author}
  {\bibfnamefont {A.}~\bibnamefont {Lucia}}, \bibinfo {author} {\bibfnamefont
  {D.}~\bibnamefont {P\'erez-Garc\'{\i}a}},\ and\ \bibinfo {author}
  {\bibfnamefont {C.}~\bibnamefont {Rouz\'e}},\ }\bibfield  {title} {\bibinfo
  {title} {Rapid thermalization of spin chain commuting hamiltonians},\ }\href
  {https://doi.org/10.1103/PhysRevLett.130.060401} {\bibfield  {journal}
  {\bibinfo  {journal} {Phys. Rev. Lett.}\ }\textbf {\bibinfo {volume} {130}},\
  \bibinfo {pages} {060401} (\bibinfo {year} {2023})}\BibitemShut {NoStop}%
\bibitem [{\citenamefont {de~Leeuw}\ \emph {et~al.}(2024)\citenamefont
  {de~Leeuw}, \citenamefont {Paletta}, \citenamefont {Pozsgay},\ and\
  \citenamefont {Vernier}}]{PhysRevB.109.054311}%
  \BibitemOpen
  \bibfield  {author} {\bibinfo {author} {\bibfnamefont {M.}~\bibnamefont
  {de~Leeuw}}, \bibinfo {author} {\bibfnamefont {C.}~\bibnamefont {Paletta}},
  \bibinfo {author} {\bibfnamefont {B.}~\bibnamefont {Pozsgay}},\ and\ \bibinfo
  {author} {\bibfnamefont {E.}~\bibnamefont {Vernier}},\ }\bibfield  {title}
  {\bibinfo {title} {Hidden quasilocal charges and gibbs ensemble in a lindblad
  system},\ }\href {https://doi.org/10.1103/PhysRevB.109.054311} {\bibfield
  {journal} {\bibinfo  {journal} {Phys. Rev. B}\ }\textbf {\bibinfo {volume}
  {109}},\ \bibinfo {pages} {054311} (\bibinfo {year} {2024})}\BibitemShut
  {NoStop}%
\bibitem [{\citenamefont {Roberts}\ \emph {et~al.}(2021)\citenamefont
  {Roberts}, \citenamefont {Lingenfelter},\ and\ \citenamefont
  {Clerk}}]{PRXQuantum.2.020336}%
  \BibitemOpen
  \bibfield  {author} {\bibinfo {author} {\bibfnamefont {D.}~\bibnamefont
  {Roberts}}, \bibinfo {author} {\bibfnamefont {A.}~\bibnamefont
  {Lingenfelter}},\ and\ \bibinfo {author} {\bibfnamefont {A.}~\bibnamefont
  {Clerk}},\ }\bibfield  {title} {\bibinfo {title} {Hidden time-reversal
  symmetry, quantum detailed balance and exact solutions of driven-dissipative
  quantum systems},\ }\href {https://doi.org/10.1103/PRXQuantum.2.020336}
  {\bibfield  {journal} {\bibinfo  {journal} {PRX Quantum}\ }\textbf {\bibinfo
  {volume} {2}},\ \bibinfo {pages} {020336} (\bibinfo {year}
  {2021})}\BibitemShut {NoStop}%
\bibitem [{\citenamefont {Roberts}\ and\ \citenamefont
  {Clerk}(2023)}]{PhysRevLett.131.190403ExactSolution}%
  \BibitemOpen
  \bibfield  {author} {\bibinfo {author} {\bibfnamefont {D.}~\bibnamefont
  {Roberts}}\ and\ \bibinfo {author} {\bibfnamefont {A.~A.}\ \bibnamefont
  {Clerk}},\ }\bibfield  {title} {\bibinfo {title} {Exact solution of the
  infinite-range dissipative transverse-field ising model},\ }\href
  {https://doi.org/10.1103/PhysRevLett.131.190403} {\bibfield  {journal}
  {\bibinfo  {journal} {Phys. Rev. Lett.}\ }\textbf {\bibinfo {volume} {131}},\
  \bibinfo {pages} {190403} (\bibinfo {year} {2023})}\BibitemShut {NoStop}%
\bibitem [{\citenamefont {Palmero}\ \emph {et~al.}(2019)\citenamefont
  {Palmero}, \citenamefont {Xu}, \citenamefont {Guo},\ and\ \citenamefont
  {Poletti}}]{PhysRevE.100.022111}%
  \BibitemOpen
  \bibfield  {author} {\bibinfo {author} {\bibfnamefont {M.}~\bibnamefont
  {Palmero}}, \bibinfo {author} {\bibfnamefont {X.}~\bibnamefont {Xu}},
  \bibinfo {author} {\bibfnamefont {C.}~\bibnamefont {Guo}},\ and\ \bibinfo
  {author} {\bibfnamefont {D.}~\bibnamefont {Poletti}},\ }\bibfield  {title}
  {\bibinfo {title} {Thermalization with detailed-balanced two-site lindblad
  dissipators},\ }\href {https://doi.org/10.1103/PhysRevE.100.022111}
  {\bibfield  {journal} {\bibinfo  {journal} {Phys. Rev. E}\ }\textbf {\bibinfo
  {volume} {100}},\ \bibinfo {pages} {022111} (\bibinfo {year}
  {2019})}\BibitemShut {NoStop}%
\bibitem [{\citenamefont {Alicki}(1976)}]{ALICKI1976249}%
  \BibitemOpen
  \bibfield  {author} {\bibinfo {author} {\bibfnamefont {R.}~\bibnamefont
  {Alicki}},\ }\bibfield  {title} {\bibinfo {title} {On the detailed balance
  condition for non-hamiltonian systems},\ }\href
  {https://doi.org/https://doi.org/10.1016/0034-4877(76)90046-X} {\bibfield
  {journal} {\bibinfo  {journal} {Reports on Mathematical Physics}\ }\textbf
  {\bibinfo {volume} {10}},\ \bibinfo {pages} {249} (\bibinfo {year}
  {1976})}\BibitemShut {NoStop}%
\bibitem [{\citenamefont {Fagnola}\ and\ \citenamefont
  {Umanit\`{a}}(2007)}]{qdbc}%
  \BibitemOpen
  \bibfield  {author} {\bibinfo {author} {\bibfnamefont {F.}~\bibnamefont
  {Fagnola}}\ and\ \bibinfo {author} {\bibfnamefont {V.}~\bibnamefont
  {Umanit\`{a}}},\ }\bibfield  {title} {\bibinfo {title} {Generators of
  detailed balance quantum markov semigroups},\ }\href
  {https://doi.org/10.1142/S0219025707002762} {\bibfield  {journal} {\bibinfo
  {journal} {Infinite Dimensional Analysis, Quantum Probability and Related
  Topics}\ }\textbf {\bibinfo {volume} {10}},\ \bibinfo {pages} {335} (\bibinfo
  {year} {2007})}\BibitemShut {NoStop}%
\bibitem [{\citenamefont {Davies}(1974)}]{Davies1974}%
  \BibitemOpen
  \bibfield  {author} {\bibinfo {author} {\bibfnamefont {E.~B.}\ \bibnamefont
  {Davies}},\ }\bibfield  {title} {\bibinfo {title} {Markovian master
  equations},\ }\href {https://doi.org/10.1007/BF01608389} {\bibfield
  {journal} {\bibinfo  {journal} {Communications in Mathematical Physics}\
  }\textbf {\bibinfo {volume} {39}},\ \bibinfo {pages} {91} (\bibinfo {year}
  {1974})}\BibitemShut {NoStop}%
\bibitem [{\citenamefont {Prosen}(2015)}]{Prosen_2015}%
  \BibitemOpen
  \bibfield  {author} {\bibinfo {author} {\bibfnamefont {T.}~\bibnamefont
  {Prosen}},\ }\bibfield  {title} {\bibinfo {title} {Matrix product solutions
  of boundary driven quantum chains},\ }\href
  {https://doi.org/10.1088/1751-8113/48/37/373001} {\bibfield  {journal}
  {\bibinfo  {journal} {Journal of Physics A: Mathematical and Theoretical}\
  }\textbf {\bibinfo {volume} {48}},\ \bibinfo {pages} {373001} (\bibinfo
  {year} {2015})}\BibitemShut {NoStop}%
\bibitem [{\citenamefont {Bu{\v{c}}a}\ and\ \citenamefont
  {Prosen}(2012)}]{Buca_2012}%
  \BibitemOpen
  \bibfield  {author} {\bibinfo {author} {\bibfnamefont {B.}~\bibnamefont
  {Bu{\v{c}}a}}\ and\ \bibinfo {author} {\bibfnamefont {T.}~\bibnamefont
  {Prosen}},\ }\bibfield  {title} {\bibinfo {title} {A note on symmetry
  reductions of the lindblad equation: transport in constrained open spin
  chains},\ }\href {https://doi.org/10.1088/1367-2630/14/7/073007} {\bibfield
  {journal} {\bibinfo  {journal} {New Journal of Physics}\ }\textbf {\bibinfo
  {volume} {14}},\ \bibinfo {pages} {073007} (\bibinfo {year}
  {2012})}\BibitemShut {NoStop}%
\bibitem [{\citenamefont {Zhang}\ \emph {et~al.}(2020)\citenamefont {Zhang},
  \citenamefont {Tindall}, \citenamefont {Mur-Petit}, \citenamefont {Jaksch},\
  and\ \citenamefont {Buča}}]{Zhang_2020}%
  \BibitemOpen
  \bibfield  {author} {\bibinfo {author} {\bibfnamefont {Z.}~\bibnamefont
  {Zhang}}, \bibinfo {author} {\bibfnamefont {J.}~\bibnamefont {Tindall}},
  \bibinfo {author} {\bibfnamefont {J.}~\bibnamefont {Mur-Petit}}, \bibinfo
  {author} {\bibfnamefont {D.}~\bibnamefont {Jaksch}},\ and\ \bibinfo {author}
  {\bibfnamefont {B.}~\bibnamefont {Buča}},\ }\bibfield  {title} {\bibinfo
  {title} {Stationary state degeneracy of open quantum systems with non-abelian
  symmetries},\ }\href {https://doi.org/10.1088/1751-8121/ab88e3} {\bibfield
  {journal} {\bibinfo  {journal} {Journal of Physics A: Mathematical and
  Theoretical}\ }\textbf {\bibinfo {volume} {53}},\ \bibinfo {pages} {215304}
  (\bibinfo {year} {2020})}\BibitemShut {NoStop}%
\bibitem [{\citenamefont {Ritort}\ and\ \citenamefont
  {Sollich}(2003)}]{doi:10.1080/0001873031000093582}%
  \BibitemOpen
  \bibfield  {author} {\bibinfo {author} {\bibfnamefont {F.}~\bibnamefont
  {Ritort}}\ and\ \bibinfo {author} {\bibfnamefont {P.}~\bibnamefont
  {Sollich}},\ }\bibfield  {title} {\bibinfo {title} {Glassy dynamics of
  kinetically constrained models},\ }\href
  {https://doi.org/10.1080/0001873031000093582} {\bibfield  {journal} {\bibinfo
   {journal} {Advances in Physics}\ }\textbf {\bibinfo {volume} {52}},\
  \bibinfo {pages} {219} (\bibinfo {year} {2003})}\BibitemShut {NoStop}%
\bibitem [{\citenamefont {Fredrickson}\ and\ \citenamefont
  {Andersen}(1984)}]{PhysRevLett.53.1244}%
  \BibitemOpen
  \bibfield  {author} {\bibinfo {author} {\bibfnamefont {G.~H.}\ \bibnamefont
  {Fredrickson}}\ and\ \bibinfo {author} {\bibfnamefont {H.~C.}\ \bibnamefont
  {Andersen}},\ }\bibfield  {title} {\bibinfo {title} {Kinetic ising model of
  the glass transition},\ }\href {https://doi.org/10.1103/PhysRevLett.53.1244}
  {\bibfield  {journal} {\bibinfo  {journal} {Phys. Rev. Lett.}\ }\textbf
  {\bibinfo {volume} {53}},\ \bibinfo {pages} {1244} (\bibinfo {year}
  {1984})}\BibitemShut {NoStop}%
\bibitem [{\citenamefont {Sala}\ \emph {et~al.}(2020)\citenamefont {Sala},
  \citenamefont {Rakovszky}, \citenamefont {Verresen}, \citenamefont {Knap},\
  and\ \citenamefont {Pollmann}}]{PhysRevX.10.011047dipole}%
  \BibitemOpen
  \bibfield  {author} {\bibinfo {author} {\bibfnamefont {P.}~\bibnamefont
  {Sala}}, \bibinfo {author} {\bibfnamefont {T.}~\bibnamefont {Rakovszky}},
  \bibinfo {author} {\bibfnamefont {R.}~\bibnamefont {Verresen}}, \bibinfo
  {author} {\bibfnamefont {M.}~\bibnamefont {Knap}},\ and\ \bibinfo {author}
  {\bibfnamefont {F.}~\bibnamefont {Pollmann}},\ }\bibfield  {title} {\bibinfo
  {title} {Ergodicity breaking arising from hilbert space fragmentation in
  dipole-conserving hamiltonians},\ }\href
  {https://doi.org/10.1103/PhysRevX.10.011047} {\bibfield  {journal} {\bibinfo
  {journal} {Phys. Rev. X}\ }\textbf {\bibinfo {volume} {10}},\ \bibinfo
  {pages} {011047} (\bibinfo {year} {2020})}\BibitemShut {NoStop}%
\bibitem [{\citenamefont {Li}\ \emph {et~al.}(2023)\citenamefont {Li},
  \citenamefont {Sala},\ and\ \citenamefont
  {Pollmann}}]{PhysRevResearch.5.043239hsfopen}%
  \BibitemOpen
  \bibfield  {author} {\bibinfo {author} {\bibfnamefont {Y.}~\bibnamefont
  {Li}}, \bibinfo {author} {\bibfnamefont {P.}~\bibnamefont {Sala}},\ and\
  \bibinfo {author} {\bibfnamefont {F.}~\bibnamefont {Pollmann}},\ }\bibfield
  {title} {\bibinfo {title} {Hilbert space fragmentation in open quantum
  systems},\ }\href {https://doi.org/10.1103/PhysRevResearch.5.043239}
  {\bibfield  {journal} {\bibinfo  {journal} {Phys. Rev. Res.}\ }\textbf
  {\bibinfo {volume} {5}},\ \bibinfo {pages} {043239} (\bibinfo {year}
  {2023})}\BibitemShut {NoStop}%
\bibitem [{\citenamefont {Balasubramanian}\ \emph {et~al.}(2024)\citenamefont
  {Balasubramanian}, \citenamefont {Gopalakrishnan}, \citenamefont
  {Khudorozhkov},\ and\ \citenamefont {Lake}}]{PhysRevX.14.021034word}%
  \BibitemOpen
  \bibfield  {author} {\bibinfo {author} {\bibfnamefont {S.}~\bibnamefont
  {Balasubramanian}}, \bibinfo {author} {\bibfnamefont {S.}~\bibnamefont
  {Gopalakrishnan}}, \bibinfo {author} {\bibfnamefont {A.}~\bibnamefont
  {Khudorozhkov}},\ and\ \bibinfo {author} {\bibfnamefont {E.}~\bibnamefont
  {Lake}},\ }\bibfield  {title} {\bibinfo {title} {Glassy word problems:
  Ultraslow relaxation, hilbert space jamming, and computational complexity},\
  }\href {https://doi.org/10.1103/PhysRevX.14.021034} {\bibfield  {journal}
  {\bibinfo  {journal} {Phys. Rev. X}\ }\textbf {\bibinfo {volume} {14}},\
  \bibinfo {pages} {021034} (\bibinfo {year} {2024})}\BibitemShut {NoStop}%
\bibitem [{\citenamefont {McDonald}\ and\ \citenamefont
  {Clerk}(2022)}]{PhysRevLett.128.033602}%
  \BibitemOpen
  \bibfield  {author} {\bibinfo {author} {\bibfnamefont {A.}~\bibnamefont
  {McDonald}}\ and\ \bibinfo {author} {\bibfnamefont {A.~A.}\ \bibnamefont
  {Clerk}},\ }\bibfield  {title} {\bibinfo {title} {Exact solutions of
  interacting dissipative systems via weak symmetries},\ }\href
  {https://doi.org/10.1103/PhysRevLett.128.033602} {\bibfield  {journal}
  {\bibinfo  {journal} {Phys. Rev. Lett.}\ }\textbf {\bibinfo {volume} {128}},\
  \bibinfo {pages} {033602} (\bibinfo {year} {2022})}\BibitemShut {NoStop}%
\bibitem [{\citenamefont {Albert}\ and\ \citenamefont
  {Jiang}(2014)}]{PhysRevA.89.022118}%
  \BibitemOpen
  \bibfield  {author} {\bibinfo {author} {\bibfnamefont {V.~V.}\ \bibnamefont
  {Albert}}\ and\ \bibinfo {author} {\bibfnamefont {L.}~\bibnamefont {Jiang}},\
  }\bibfield  {title} {\bibinfo {title} {Symmetries and conserved quantities in
  lindblad master equations},\ }\href
  {https://doi.org/10.1103/PhysRevA.89.022118} {\bibfield  {journal} {\bibinfo
  {journal} {Phys. Rev. A}\ }\textbf {\bibinfo {volume} {89}},\ \bibinfo
  {pages} {022118} (\bibinfo {year} {2014})}\BibitemShut {NoStop}%
\bibitem [{\citenamefont {Li}\ \emph {et~al.}(2024)\citenamefont {Li},
  \citenamefont {Pollmann}, \citenamefont {Read},\ and\ \citenamefont
  {Sala}}]{li2024highlyentangled}%
  \BibitemOpen
  \bibfield  {author} {\bibinfo {author} {\bibfnamefont {Y.}~\bibnamefont
  {Li}}, \bibinfo {author} {\bibfnamefont {F.}~\bibnamefont {Pollmann}},
  \bibinfo {author} {\bibfnamefont {N.}~\bibnamefont {Read}},\ and\ \bibinfo
  {author} {\bibfnamefont {P.}~\bibnamefont {Sala}},\ }\href@noop {} {\bibinfo
  {title} {Highly-entangled stationary states from strong symmetries}}
  (\bibinfo {year} {2024}),\ \Eprint {https://arxiv.org/abs/2406.08567}
  {arXiv:2406.08567 [quant-ph]} \BibitemShut {NoStop}%
\bibitem [{\citenamefont {Evans}(1977)}]{evans1977irreducible}%
  \BibitemOpen
  \bibfield  {author} {\bibinfo {author} {\bibfnamefont {D.~E.}\ \bibnamefont
  {Evans}},\ }\bibfield  {title} {\bibinfo {title} {Irreducible quantum
  dynamical semigroups},\ }\href
  {https://doi.org/https://doi.org/10.1007/BF01614091} {\bibfield  {journal}
  {\bibinfo  {journal} {Commun.Math. Phys}\ }\textbf {\bibinfo {volume} {54}},\
  \bibinfo {pages} {293–297} (\bibinfo {year} {1977})}\BibitemShut {NoStop}%
\bibitem [{\citenamefont {Frigerio}(1977)}]{Frigerio1977}%
  \BibitemOpen
  \bibfield  {author} {\bibinfo {author} {\bibfnamefont {A.}~\bibnamefont
  {Frigerio}},\ }\bibfield  {title} {\bibinfo {title} {Quantum dynamical
  semigroups and approach to equilibrium},\ }\href
  {https://doi.org/10.1007/BF00398571} {\bibfield  {journal} {\bibinfo
  {journal} {Letters in Mathematical Physics}\ }\textbf {\bibinfo {volume}
  {2}},\ \bibinfo {pages} {79} (\bibinfo {year} {1977})}\BibitemShut {NoStop}%
\bibitem [{\citenamefont {Baumgartner}\ and\ \citenamefont
  {Narnhofer}(2008)}]{Baumgartner_2008}%
  \BibitemOpen
  \bibfield  {author} {\bibinfo {author} {\bibfnamefont {B.}~\bibnamefont
  {Baumgartner}}\ and\ \bibinfo {author} {\bibfnamefont {H.}~\bibnamefont
  {Narnhofer}},\ }\bibfield  {title} {\bibinfo {title} {Analysis of quantum
  semigroups with gks–lindblad generators:
  {\uppercase\expandafter{\romannumeral2}}. general},\ }\href
  {https://doi.org/10.1088/1751-8113/41/39/395303} {\bibfield  {journal}
  {\bibinfo  {journal} {Journal of Physics A: Mathematical and Theoretical}\
  }\textbf {\bibinfo {volume} {41}},\ \bibinfo {pages} {395303} (\bibinfo
  {year} {2008})}\BibitemShut {NoStop}%
\bibitem [{\citenamefont {Zhang}\ and\ \citenamefont
  {Barthel}(2024)}]{Zhang_2024}%
  \BibitemOpen
  \bibfield  {author} {\bibinfo {author} {\bibfnamefont {Y.}~\bibnamefont
  {Zhang}}\ and\ \bibinfo {author} {\bibfnamefont {T.}~\bibnamefont
  {Barthel}},\ }\bibfield  {title} {\bibinfo {title} {Criteria for davies
  irreducibility of markovian quantum dynamics},\ }\href
  {https://doi.org/10.1088/1751-8121/ad2a1e} {\bibfield  {journal} {\bibinfo
  {journal} {Journal of Physics A: Mathematical and Theoretical}\ }\textbf
  {\bibinfo {volume} {57}},\ \bibinfo {pages} {115301} (\bibinfo {year}
  {2024})}\BibitemShut {NoStop}%
\bibitem [{\citenamefont
  {Landsman}(1998)}]{landsman1998lecturenotescalgebrashilbert}%
  \BibitemOpen
  \bibfield  {author} {\bibinfo {author} {\bibfnamefont {N.~P.}\ \bibnamefont
  {Landsman}},\ }\href {https://arxiv.org/abs/math-ph/9807030} {\bibinfo
  {title} {Lecture notes on c*-algebras, hilbert c*-modules, and quantum
  mechanics}} (\bibinfo {year} {1998}),\ \Eprint
  {https://arxiv.org/abs/math-ph/9807030} {arXiv:math-ph/9807030} \BibitemShut
  {NoStop}%
\bibitem [{\citenamefont {\ifmmode \check{Z}\else
  \v{Z}\fi{}nidari\ifmmode~\check{c}\else
  \v{c}\fi{}}(2011)}]{PhysRevE.83.011108}%
  \BibitemOpen
  \bibfield  {author} {\bibinfo {author} {\bibfnamefont {M.}~\bibnamefont
  {\ifmmode \check{Z}\else \v{Z}\fi{}nidari\ifmmode~\check{c}\else
  \v{c}\fi{}}},\ }\bibfield  {title} {\bibinfo {title} {Solvable quantum
  nonequilibrium model exhibiting a phase transition and a matrix product
  representation},\ }\href {https://doi.org/10.1103/PhysRevE.83.011108}
  {\bibfield  {journal} {\bibinfo  {journal} {Phys. Rev. E}\ }\textbf {\bibinfo
  {volume} {83}},\ \bibinfo {pages} {011108} (\bibinfo {year}
  {2011})}\BibitemShut {NoStop}%
\bibitem [{\citenamefont {Salberger}\ and\ \citenamefont
  {Korepin}(2017)}]{doi:10.1142/S0129055X17500313fredkinchain}%
  \BibitemOpen
  \bibfield  {author} {\bibinfo {author} {\bibfnamefont {O.}~\bibnamefont
  {Salberger}}\ and\ \bibinfo {author} {\bibfnamefont {V.}~\bibnamefont
  {Korepin}},\ }\bibfield  {title} {\bibinfo {title} {Entangled spin chain},\
  }\href {https://doi.org/10.1142/S0129055X17500313} {\bibfield  {journal}
  {\bibinfo  {journal} {Reviews in Mathematical Physics}\ }\textbf {\bibinfo
  {volume} {29}},\ \bibinfo {pages} {1750031} (\bibinfo {year}
  {2017})}\BibitemShut {NoStop}%
\bibitem [{app()}]{appendix_note}%
  \BibitemOpen
  \href@noop {} {}\bibinfo {note} {See Supplemental Material for additional
  results of Fredkin model under boundary driving and more examples of other
  models.}\BibitemShut {Stop}%
\bibitem [{\citenamefont {\ifmmode \check{Z}\else
  \v{Z}\fi{}nidari\ifmmode~\check{c}\else
  \v{c}\fi{}}(2015)}]{PhysRevE.92.042143}%
  \BibitemOpen
  \bibfield  {author} {\bibinfo {author} {\bibfnamefont {M.}~\bibnamefont
  {\ifmmode \check{Z}\else \v{Z}\fi{}nidari\ifmmode~\check{c}\else
  \v{c}\fi{}}},\ }\bibfield  {title} {\bibinfo {title} {Relaxation times of
  dissipative many-body quantum systems},\ }\href
  {https://doi.org/10.1103/PhysRevE.92.042143} {\bibfield  {journal} {\bibinfo
  {journal} {Phys. Rev. E}\ }\textbf {\bibinfo {volume} {92}},\ \bibinfo
  {pages} {042143} (\bibinfo {year} {2015})}\BibitemShut {NoStop}%
\bibitem [{\citenamefont {Han}\ \emph {et~al.}(2024)\citenamefont {Han},
  \citenamefont {Chen},\ and\ \citenamefont {Lake}}]{han2024exponentially}%
  \BibitemOpen
  \bibfield  {author} {\bibinfo {author} {\bibfnamefont {Y.}~\bibnamefont
  {Han}}, \bibinfo {author} {\bibfnamefont {X.}~\bibnamefont {Chen}},\ and\
  \bibinfo {author} {\bibfnamefont {E.}~\bibnamefont {Lake}},\ }\href@noop {}
  {\bibinfo {title} {Exponentially slow thermalization and the robustness of
  hilbert space fragmentation}} (\bibinfo {year} {2024}),\ \Eprint
  {https://arxiv.org/abs/2401.11294} {arXiv:2401.11294 [quant-ph]} \BibitemShut
  {NoStop}%
\bibitem [{\citenamefont {Ness}(2013)}]{PhysRevE.88.022121}%
  \BibitemOpen
  \bibfield  {author} {\bibinfo {author} {\bibfnamefont {H.}~\bibnamefont
  {Ness}},\ }\bibfield  {title} {\bibinfo {title} {Nonequilibrium density
  matrix for quantum transport: Hershfield approach as a mclennan-zubarev form
  of the statistical operator},\ }\href
  {https://doi.org/10.1103/PhysRevE.88.022121} {\bibfield  {journal} {\bibinfo
  {journal} {Phys. Rev. E}\ }\textbf {\bibinfo {volume} {88}},\ \bibinfo
  {pages} {022121} (\bibinfo {year} {2013})}\BibitemShut {NoStop}%
\bibitem [{\citenamefont {Subotnik}\ \emph {et~al.}(2009)\citenamefont
  {Subotnik}, \citenamefont {Hansen}, \citenamefont {Ratner},\ and\
  \citenamefont {Nitzan}}]{10.1063/1.3109898}%
  \BibitemOpen
  \bibfield  {author} {\bibinfo {author} {\bibfnamefont {J.~E.}\ \bibnamefont
  {Subotnik}}, \bibinfo {author} {\bibfnamefont {T.}~\bibnamefont {Hansen}},
  \bibinfo {author} {\bibfnamefont {M.~A.}\ \bibnamefont {Ratner}},\ and\
  \bibinfo {author} {\bibfnamefont {A.}~\bibnamefont {Nitzan}},\ }\bibfield
  {title} {\bibinfo {title} {{Nonequilibrium steady state transport via the
  reduced density matrix operator}},\ }\href
  {https://doi.org/10.1063/1.3109898} {\bibfield  {journal} {\bibinfo
  {journal} {The Journal of Chemical Physics}\ }\textbf {\bibinfo {volume}
  {130}},\ \bibinfo {pages} {144105} (\bibinfo {year} {2009})}\BibitemShut
  {NoStop}%
\end{thebibliography}%


\end{document}